\documentclass[onecolumn,showpacs,preprintnumbers,nofootinbib,amssymb]{revtex4-1}

\usepackage[linktocpage]{hyperref}
\usepackage{graphicx, epsfig,amsmath}
\usepackage{url}
\usepackage{wrapfloat}

\usepackage[usenames]{color}

\def\Lie{\mathcal{L}}
\def\D{\mathcal{D}}

\def\H{\mathcal{H}}
\def\M{\mathcal{M}}
\def\N{\mathcal{N}}
\def\O{\mathcal{O}}
\def\P{\mathcal{P}}

\def\S{\mathcal{S}}
\def\T{\mathcal{T}}

\def\bk{\bar{k}}
\def\bl{\bar{l}}
\def\bm{\bar{m}}
\def\bn{\bar{n}}
\def\bp{\bar{p}}
\def\bq{\bar{q}}
\def\bmu{\bar{\mu}}
\def\bnu{\bar{\nu}}
\def\brho{\bar{\rho}}
\def\bsigma{\bar{\sigma}}
\def\tA{\tilde{A}}
\def\tD{\tilde{D}}
\def\tg{\tilde{\gamma}}
\def\tG{\tilde{\Gamma}}
\def\tR{\tilde{R}}

\def\p{\partial}

\begin{document}


%
%

\title{Lecture Notes:
\\
NUMERICAL RELATIVITY IN HIGHER DIMENSIONAL SPACETIMES}

\author{Helvi Witek}\email{H.Witek@damtp.cam.ac.uk}
\affiliation{
Department of Applied Mathematics and Theoretical Physics,
Centre for Mathematical Sciences, University of Cambridge,
Wilberforce Road, Cambridge CB3 0WA, UK
}


\begin{abstract}
Black holes are among the most exciting phenomena predicted by General Relativity
and play a key role in fundamental physics.
Many interesting phenomena involve dynamical black hole configurations in the high curvature regime
of gravity.
In these lecture notes I will summarize the main numerical relativity techniques 
to explore highly dynamical phenomena, such as black hole collisions,
in generic $D$-dimensional spacetimes.
The present notes are based on my lectures given at the
NR/HEP2 spring school at IST/ Lisbon (Portugal) from March 11 -- 14, 2013.

\end{abstract}

%

\maketitle
\tableofcontents
\clearpage
\newpage
\section{Introduction} \label{sec:Intro}
Black holes (BHs) are among the most exciting objects predicted by General Relativity (GR)  
-- our most beloved theory of gravity to-date. 
Nowadays, BHs have outgrown their status as mere exotic mathematical constructions and
there is compelling observational evidence for their existence: 
The trajectories of stars close to the centre of the Milky Way hint at the presence of 
a supermassive BH (SMBH) with $M\sim4.2\cdot10^{6} M_{\odot}$ 
and, in fact, SMBHs with $M\sim10^{6}-10^{9}M_{\odot}$ are expected to be at the centre of most 
galaxies~\cite{Rees:1984si,Begelman:1980vb,Ferrarese:2004qr}.
Their ``light'' counterparts with a few solar masses $M\sim3-30M_{\odot}$
are conjectured to make up a large part of the galaxies' 
population~\cite{McClintock:2009as,Antoniadis:2013pzd,Seoane:2013qna}.
However, the importance of understanding the physics of BHs 
goes far beyond their role in astrophysics.
In fact, BHs are expected to be key players in a wide range of 
fundamental theories,
including astrophysics and cosmology, (modified) gravity theories, high energy physics and 
the gauge/gravity duality.
In a recent review Cardoso et al.~\cite{Cardoso:2012qm}
outlined the exciting new physics awaiting us.
Exploring BH phenomena in GR cooks down to 
investigating gravity in four or higher dimensional 
spacetimes with generic asymptotics  
described by Einstein's equations
\begin{align}
\label{eq:EEs}
R_{MN} - \frac{1}{2} g_{MN} R + g_{MN} \Lambda = & 8\pi G_{D} T_{MN}
\,,
\end{align}
where $\Lambda$ is the cosmological constant~\footnote{$\Lambda < 0$ 
corresponds to (asymptotically) anti-de Sitter spacetimes, while
$\Lambda>0$ correponds to (asymptotically) de-Sitter spacetimes.}
and $G_{D}$ denotes the $D$-dimensional Newton constant.
Despite the apparently simple form of Eq.~\eqref{eq:EEs} they are, 
in fact, a set of $D(D+1)/2$ coupled, non-linear 
partial differential equations (PDEs) of mixed elliptic, hyperbolic and parabolic type,
which in general are non-separable and hard to solve.

Depending on the particular task at hand there are different solution techniques available,
some of which are described in this collection of lecture notes.
For expample, Rostworoski presents a treatment of asymptotically anti-de Sitter spacetimes in spherical
symmetry~\cite{Maliborski:2013via}.
Instead, Pani~\cite{Pani:2013pma} as well as Sampaio~\cite{Sampaio:2013faa} in their contributions to the 
lecture notes focus on perturbative treatments of the equations of motion (EOMs). 
However, for highly dynamical systems involving strong fields
perturbative methods would break down and we have to solve the full set 
of Einstein's equations using Numerical Relativity (NR) methods.
For this purpose the EoMs are typically rewritten as a 
Cauchy problem, such that they become a set of hyperbolic (or time evolution) equations
together with constraint equations of elliptic type.
Then, in a so-called \textit{free evolution scheme}, the constraints are solved for the initial data.
Because of the Bianchi identities, in the continuum limit 
the constraints are satisfied throughout the time evolution if they have been 
fulfilled initially. Therefore, instead of solving for the constraints on
each timeslice it is sufficient to check them during a simulation.
Solutions to the initial value problem and the construction of initial data
applied to higher dimensional spacetimes
is discussed in Okawa's contribution to these lecture notes~\cite{Okawa:2013afa}.

One of the key ingredients for a successful numerical scheme is the particular 
formulation of Einstein's equations as Cauchy problem.
A necessary condition for numerical stability is the well-posedness of the continuum PDE system
as is discussed in Hilditch's contribution to these lecture notes~\cite{Hilditch:2013sba}.
Typically, NR methods imply heavy numerical simulations in $3+1$-dimensional setups.
Implementing such a scheme with various (highly involved) numerical techniques, such as adaptive mesh-refinement, parallelization
of the code, etc.
is a huge effort. In their contribution to these lecture notes
Zilh\~{a}o \& L\"{o}ffler~\cite{Zilhao:2013hia} introduce the publicly available 
\textit{Einstein toolkit}~\cite{Loffler:2011ay,EinsteinToolkit}
a code developed by many groups in the NR community and specifically designed to solve Einstein's equations
on supercomputers.
Finally, Almeida in his contibution~\cite{Sergio} discusses how a NR implementation 
has to be developed for efficient High Performance Computing.

Instead, I will focus on the evolution sector 
with the spotlight on higher dimensional gravity.
In particular, I will introduce the $(D-1)+1$ splitting of spacetime and the decomposition
of Einstein's equations which is the basis for a Cauchy formulation
in Sec.~\ref{sec:Decomp}.
As one example of a well-posed formulation of Einstein's equations I will present the
widely used
Baumgarte-Shapiro-Shibata-Nakamura formalism~\cite{Shibata:1995we,Baumgarte:1998te} (BSSN)
and refer the interested reader to 
Refs.~\cite{Garfinkle:2001ni,Pretorius:2005gq,Pretorius:2006tp,
Pretorius:2004jg,Lindblom:2005qh,Szilagyi:2009qz,Lehner:2011wc}.
for the alternative Generalized Harmonic formulation (GHG)
and 
Refs.~\cite{Bona:2003qn,Alic:2008pw,Bernuzzi:2009ex, Weyhausen:2011cg, Cao:2011fu, Hilditch:2012fp}
for the lately developed Z4c formulation 
and Refs.~\cite{Alic:2011gg,Alic:2013xsa}
for its covariant counterpart CCZ4
which wed the advantages of both the BSSN and GHG schemes.

While these ingredients are well known for $4$-dimensional spacetimes
their generalization to higher dimensional spacetimes requires more work.
In order to be feasible for currently available computational resources,
any numerical scheme should be effectively $3+1$ dimensional or less.
Therefore, I present two independent schemes providing this reduction,
namely the \textit{Cartoon method} 
in Sec.~\ref{sec:Cartoon}
and a formalism based on the \textit{dimensional reduction by isometry}
in Sec.~\ref{sec:DimRed}.

\newpage
This chapter is based on the lectures that I gave at the NR/HEP2 spring school at the
Instituto Superior T\'{e}cnico in Lisbon/ Portugal from
March 11 -- 14, 2013.
The material corresponding to these lecture notes, such as \textsc{mathematica} notebooks,
animations and slides are available at Ref.~\cite{ConfWeb}.
Along with the notes I provide exercises in each section and
solutions will be given in~\ref{asec:app1} --~\ref{asec:app3}.
Unless denoted otherwise I will use 
geometric units $c=G=1$ and 
employ the following notation:

\begin{tabular}{lll}
$M, N, \ldots$     & $= 0,\ldots,(D-1)$ & for $D$-dimensional spacetime indices, \\
$\bm,\bn,\ldots$   & $= 1,\ldots,(D-1)$ & for $(D-1)$-dimensional spatial indices, \\
$\mu,\nu,\ldots$   & $= 0,\ldots,3$     & for $4$-dimensional spacetime indices, \\ 
$i,j,\ldots$       & $= 1,2,3$          & for $3$-dimensional spatial indices,\\
$\bmu,\bnu,\ldots$ & $= 4,\ldots,(D-1)$ & for extra-dimensional spatial indices.\\
\end{tabular}

\section{$(D-1)+1$ formulation of Einsteins equations}\label{sec:Decomp}
Most NR schemes rely on the formulation of Einstein's equations
as time evolution or Cauchy problem. 
Because this implies re-writing Eq.~\eqref{eq:EEs} as first order in time/ second order in space PDEs,
I here summarize the main aspects of the $(D-1)+1$ decomposition of spacetime and of Einstein's equations.
For a more detailed discussion I 
refer the reader to textbooks and reviews of NR in $4$-dimensional asymptotically flat 
spacetimes~\cite{Alcubierre:2008,York:1979,Gourgoulhon:2007ue,Pretorius:2007nq,
Centrella:2010mx,Hinder:2010vn,Sperhake:2011xk,Baumgarte:2002jm,Baumgarte2010}
and to Refs.~\cite{Zilhao:2013gu,Witek:2013koa,Yoshino:2011zz,Yoshino:2011zza,Sperhake:2013qa,
Lehner:2011wc}
for its extension to higher dimensional spacetimes.
%
\subsection{$(D-1)+1$ decomposition of spacetime}\label{ssec:DecompST}
\begin{figure}[htpb!]
\begin{center}
\includegraphics[width=0.5\textwidth]{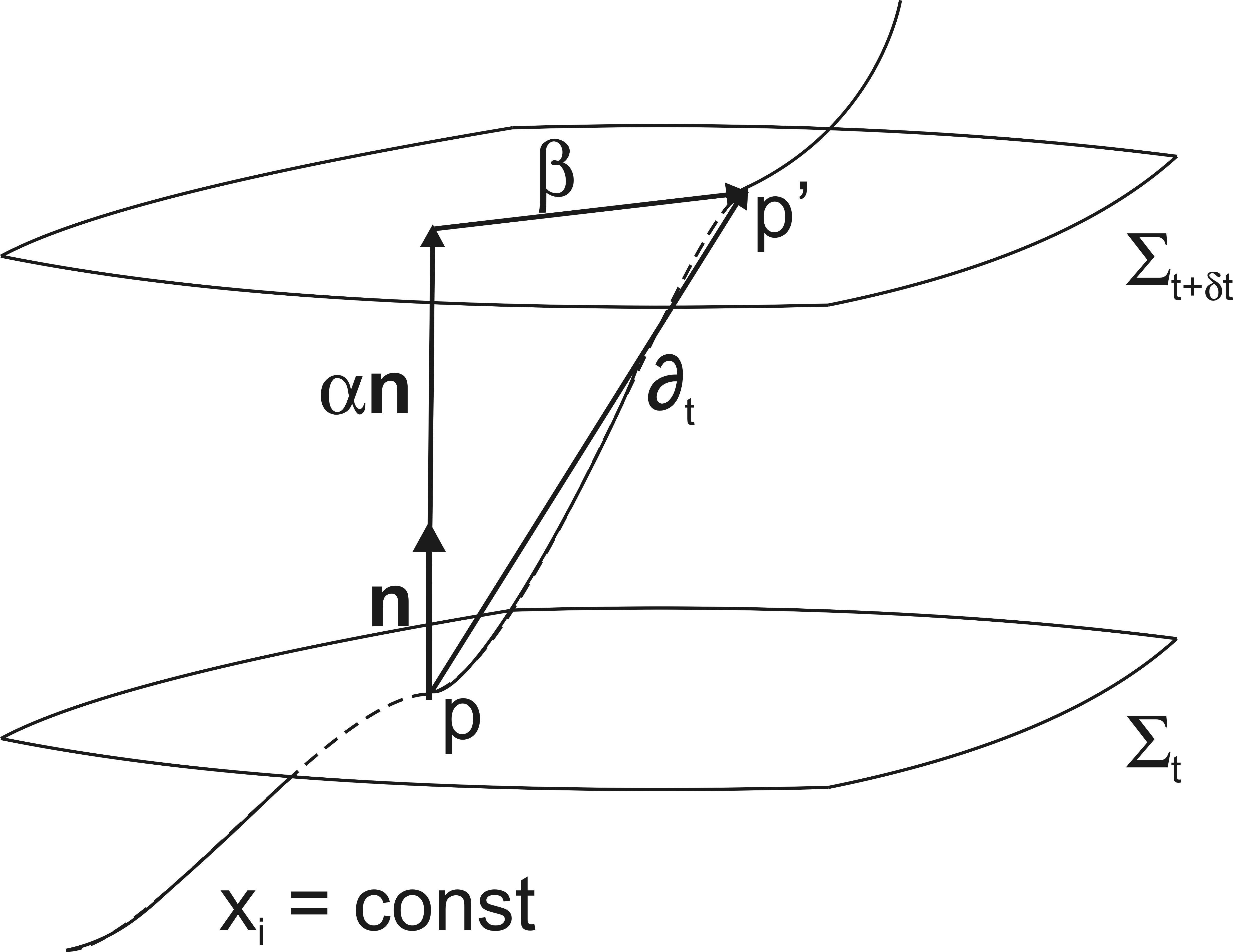}
\vspace*{8pt}
\caption{
\label{fig:foliation}
Illustration of the foliation of spacetime into 
spatial hypersurfaces $(\Sigma_{t},\gamma_{\bm\bn})$
labelled by the time parameter $t$. The coordinates are described by the lapse function $\alpha$
and the shift vector $\beta^{\bm}$.  
Taken from Witek~\cite{Witek:2013koa}.
}
\end{center}
\end{figure}

\noindent{{\bf{Foliation of the spacetime:}}} 
At the core of most NR schemes lies the decomposition of a $D$-dimensional spacetime $(\M, g_{MN})$ 
into $(D-1)$-dimensional spatial hypersurfaces $(\Sigma_t, \gamma_{\bm\bn})$ which are labelled 
by a time parameter $t\in\mathbb{R}$
\begin{align}
\,^{(D)}\M = & \,^{(D-1)}\Sigma_{t} \times \mathbb{R}
\end{align}
as illustrated in Fig.~\ref{fig:foliation}.
The $D$-dimensional spacetime $(\M, g_{MN})$ is foliated 
into a stack of spacelike hypersurfaces $(\Sigma_t, \gamma_{\bm\bn})$ of co-dimension one,
such that the vector $n^{M}$ normal to a hypersurface is timelike,
i.e., $n_M n^{M} = -1$.
The $D$-dimensional spacetime metric $g_{MN}$ and the induced,
$(D-1)$-dimensional spatial metric $\gamma_{\bm\bn}$ are related via
\begin{align}
\label{eq:GamG1}
\gamma_{MN} = g_{MN} + n_M n_N\,,& \quad
\gamma^{MN} = g^{MN} + n^M n^N
\,.
\end{align}
It is straightforward to show that $\gamma_{\bm\bn}$ is indeed spatial, i.e., $\gamma_{MN} n^{M}=0$.
In terms of the induced metric $\gamma_{\bm\bn}$, the spacetime geometry can
now be described by the line element
\begin{align}
\label{eq:DecompLE}
ds^2 = & g_{MN} dx^M dx^N 
     =   -   \left( \alpha^2 - \beta_{\bm} \beta^{\bm} \right) dt^2
         + 2 \gamma_{\bm\bn} \beta^{\bm} dt dx^{\bn}
         +   \gamma_{\bm\bn} dx^{\bm} dx^{\bn}
\,.
\end{align}
The function $\alpha$ is called the \textit{lapse} function, and measures how much proper time has elapsed 
between two timeslices $\,^{(D-1)}\Sigma_t$ and $^{(D-1)}\Sigma_{t+\delta t}$. 
In other words, the lapse 
relates the coordinate time $t$ to the time measured by an 
Eulerian observer~\footnote{An Eulerian observer is an observer moving along the normal vector, i.e.,
orthogonal to the hypersuface.}.
Instead, the \textit{shift} vector $\beta^{\bm}$ indicates by how much the spatial coordinates 
of a point $p'\in\,^{(D-1)}\Sigma_{t +\delta t}$ are shifted or displaced 
as compared to the point in $\,^{(D-1)}\Sigma_{t +\delta t}$ obtained from 
going just along the normal vector starting from the (original) point $p\in^{(D-1)}\Sigma_t$.
In more technical terms, the shift vector measures the relative velocity between an Eulerian observer and 
lines of constant coordinates.
In particular, the shift vector is purely spatial $\beta^M=(0,\beta^{\bm})$.
Together the lapse and shift $(\alpha, \beta^{\bm})$ encode the coordinate degrees of freedom in gravity
and are often referred to as gauge variables.
As depicted in Fig.~\ref{fig:foliation}, a vector $t^M$ pointing from a 
point $p\in^{(D-1)}\Sigma$ to a point $p'\in^{(D-1)}\Sigma_{t+\delta t}$ 
is given by
\begin{align}
\label{eq:talpbet}
t^M = & \alpha n^M + \beta^M
\,.
\end{align}
In terms of the gauge variables $(\alpha,\beta^{\bm})$ the normal vector $n^M$ can be expressed as 
\begin{align}
n_M = -\alpha\left(1,0,...,0\right)\,,&\quad n^M = \frac{1}{\alpha}\left(1,-\beta^{\bm}\right)
\,.
\end{align}

\noindent{{\bf{Decomposition of tensors:}}} 
The relation~\eqref{eq:GamG1} between the spacetime metric $g_{MN}$ and the induced metric $\gamma_{\bm\bn}$
defines the projection operator
\begin{align}
\label{eq:projop}
\perp=\gamma^{M}{}_{N} = & \delta^{M}{}_{N} + n^{M} n_{N}
\,.
\end{align}
By employing the projection operator $\perp$ and normal vector $n^M$
any $(p,q)$-tensor $T^{M_1...M_p}{}_{N_1...N_q}\in\M$~\footnote{ 
Strictly speaking $T^{M_1...M_p}{}_{N_1...N_q}$ are components of a tensor 
with respect to the basis in the tangent and cotangent spaces $\mathcal{T}_p(\mathcal{M})$
and $\mathcal{T}^{\ast}{}_{p}(\mathcal{M})$ at a point $p\in\mathcal{M}$.
However, now and in the reminder of these lecture notes I will use this abbreviated notation
which really means 
$T\in \mathcal{T}_{p}(\mathcal{M})\times\ldots\times\mathcal{T}_{p}(\mathcal{M}) \times
\mathcal{T}^{\ast}_{p}(\mathcal{M})\times\ldots\times\mathcal{T}^{\ast}{}_{p}(\mathcal{M})$.}
can be decomposed into
its normal, purely spatial and mixed components. In the following I will illustrate this 
decomposition exemplarily for a rank-$2$ tensor $T_{MN}\in\M$ 
which can be generalized in a straightforward manner. 
Let us denote the normal component by $\N$, the purely spatial component by $\S_{MN}$ and
the mixed component by $\T_{M}$ with
\begin{align}
\label{eq:TensProj}
\N      = & T_{MN} n^M n^N 
\,,\qquad
\T_M    = - \gamma^{K}{}_{M} T_{KN} n^N 
\,,\qquad
\S_{MN} =   \gamma^{K}{}_{M} \gamma^{L}{}_{N} T_{KL}
\,.
\end{align}
Then, the $D$-dimensional spacetime tensor $T_{MN}\in\M$ is reconstructed from
\begin{align}
T_{MN} = & \S_{MN} + \T_{M} n_{N} + \T_{N} n_{M} + \N n_{M} n_{N} 
\,.
\end{align}
Let us denote the covariant derivative with respect to the spatial metric $\gamma_{\bm\bn}$ by $D_{\bm}$.
The (spatial) covariant derivative $D_{\bm}$ of any $(p,q)$-tensor $T^{M_1...M_p}{}_{N_1...N_q} \in\,^{(D-1)}\Sigma$
is related to the covariant derivative $\nabla_M$ associated with the spacetime metric $g_{MN}$
through
\begin{align}
D_{\bk} T^{\bm_1...\bm_p}{}_{\bn_1...\bn_q} = &
\gamma^{\bm_1}{}_{M_1}...\gamma^{\bm_p}{}_{M_p}\gamma^{N_1}{}_{\bn_1}...\gamma^{N_q}{}_{\bn_q} \gamma^{K}{}_{\bk}
\nabla_K T^{M_1...M_p}{}_{N_1...N_q}
\,.
\end{align}
Let us further note, that the metric compatible, torsion-free connection coefficients
with respect to the spatial metric $\gamma_{\bm\bn}$ are computed with
\begin{align}
\Gamma^{\bk}{}_{\bm\bn} = & \frac{1}{2} \gamma^{\bk\bl}
                            \left(\p_{\bm}\gamma_{\bl\bn} + \p_{\bn}\gamma_{\bm\bl} - \p_{\bl} \gamma_{\bm\bn} \right)
\,.
\end{align}

\begin{wrapfigure}[20]{l}{0.60\textwidth}
\begin{center}
\includegraphics[width=0.35\textwidth]{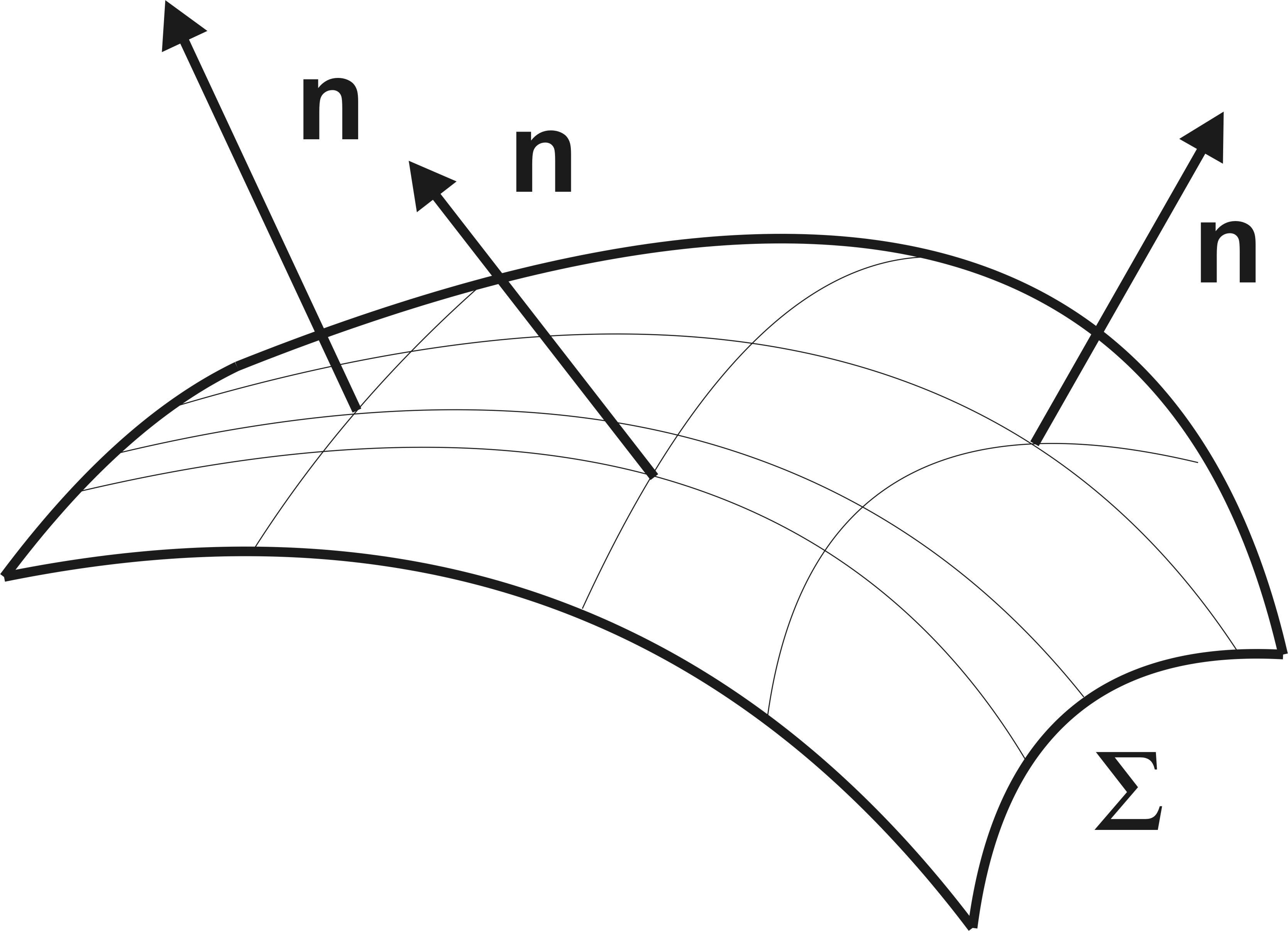}
\end{center}
\vspace*{8pt}
\caption{ \label{fig:extrcurv}
Illustration of the extrinsic curvature of a spatial hypersurface $\Sigma_t$.
Taken from Witek~\cite{Witek:2013koa}. }
\end{wrapfigure}

\noindent{{\bf{Extrinsic curvature:}}} 
So far I have focused on the description of coordinates and the induced metric 
on a spatial hypersurface $\,^{(D-1)}\Sigma_t$. 
In order to fully describe the entire spacetime, we also have to charaterize how a
hypersurface is embedded into the spacetime manifold $\,^{(D)}\M$. 
We accomplish this task by introducing the \textit{extrinsic curvature} $K_{\bm\bn}$.
Geometrically, the extrinsic curvature is a measure of how the direction of the normal vector $n^M$ changes 
as it is transported along a timeslice, as depicted in Fig.~\ref{fig:extrcurv}.
In more formal terms, the extrinsic curvature is then defined as the (projected)
covariant derivative of the normal vector
\begin{align}
\label{eq:def1Kmn}
K_{MN} = & -\gamma^{K}{}_{M}\nabla_K n_{N} = - \nabla_M n_N - n_M a_N
\,,
\end{align}
where 
$a_M = n^{N}\nabla_N n_M = \tfrac{1}{\alpha} D_M \alpha$
is the acceleration of an observer traveling along the normal vector. 
Note, that the definition~\eqref{eq:def1Kmn} relies solely on the geometry of the spacetime
and that the extrinsic curvature is symmetric and purely spatial.
The latter property implies $K_{00}=K_{0i}=0$ in coordinates adapted to the spacetime decomposition
and therefore we will typically only use the spatial components $K_{MN}=K_{\bm\bn}$.
Besides this nice geometical interpretation of the extrinsic curvature 
it can also be viewed as a kinematical degree of freedom:
In the presence of a stack or \textit{foliation} of the spacelike slices
(as is the case in our approach), 
we can relate the extrinsic curvature to the Lie derivative of the spatial metric $\gamma_{\bm\bn}$
along the normal vector $n^M$
\begin{align}
\label{eq:def2Kmn}
K_{\bm\bn} = & -\frac{1}{2\alpha} \Lie_{\alpha n} \gamma_{\bm\bn} 
           =   -\frac{1}{2\alpha} \left( \p_t - \Lie_{\beta}\right) \gamma_{\bm\bn}
\,.
\end{align}

This relation provides the kinematical interpretation of the extrinsic curvature as ``momentum''
or ``time derivative'' of the induced metric $\gamma_{\bm\bn}$
as measured by an Eulerian observer.
So far, all quantities have been derived from purely geometrical concepts,
and therefore merely allow for kinematical descriptions.
The dynamics will enter the game only through the Einstein's equations as I will discuss in the following section.
\vspace{2.0cm}

\subsection{$(D-1)+1$ decomposition of Einstein's equations}\label{ssec:DecompEE}
In the previous section I have focused on purely geometrical concepts,
providing us with only kinematical degrees of freedom.
In order to grasp the dynamical degrees of freedom we need to solve
the EoMs. 
Therefore, the next step is the $(D-1)+1$ splitting of Einstein's equations~\eqref{eq:EEs}.

\noindent{{\bf{Decomposition of the Riemann tensor and the Gauss-Codazzi relations:}}}
As preparation for this task we first focus on the $(D-1)+1$ decomposition of the 
Riemann tensor which will yield the \textit{Gauss-Codazzi} equations.
First, let us recall 
that the Riemann tensor measures the non-commutativity between two succesive covariant derivatives
giving the Ricci identity
\begin{subequations}
\label{eq:RicciIDs}
\begin{align}
\label{eq:RicciID1}
\left(\nabla_{M}\nabla_{N}-\nabla_{N}\nabla_{M}\right)V^L = \,^{(D)}R^{K}{}_{LMN} V^L\,,&\quad V^L\in\M
\,,\\
\label{eq:RicciID2}
\left(D_{\bm}D_{\bn} - D_{\bn}D_{\bm} \right) v^{\bl} = \,^{(D-1)}R^{\bk}{}_{\bl\bm\bn} v^{\bl}\,,&\quad v^{\bl}\in\Sigma_t
\end{align}
\end{subequations}
for the Riemann tensors $\,^{(D)}R^{K}{}_{LMN}\in\,^{(D)}\M$
and $\,^{(D-1)}R^{\bk}{}_{\bl\bm\bn} \in\,^{(D-1)}\Sigma_t$,
associated, respectively, with the spacetime metric $g_{MN}$ and spatial metric $\gamma_{\bm\bn}$.
Along the way towards the Gauss-Codazzi equations we will need the relation 
\begin{align}
\label{eq:DDv}
D_M D_N v_L = & \gamma^{A}{}_{M} \gamma^{B}{}_{N} \gamma^{C}{}_{L} \nabla_A \nabla_B v_C
                - K_{MN} \gamma^{C}{}_{L} n^D \nabla_D v_C
                - K_{ML} K^{D}{}_{N} v_D
\,,
\end{align}
for a spatial vector field $v_{\bl}$, 
where I have used 
\begin{align}
\nabla_M \gamma^{A}{}_{B} = &- n^A K_{MB} - n_B K^{A}{}_{M} - n^A n_M a_B - n_B n_M a^A 
\,.
\end{align}
Now we insert Eq.~\eqref{eq:DDv} into the Ricci identities~\eqref{eq:RicciIDs}
and consider the various projections of the Riemann tensor.
The only non-trivial components, as can be seen from the symmetries of the Riemann tensor,
yield the \textit{Gauss-Codazzi} relations
(see e.g., Refs.~\cite{Alcubierre:2008,York:1979,Wald:1984rg})
\begin{subequations}
\label{eq:GaussCodazzi}
\begin{align}
\gamma^{L}{}_{A}\gamma^{B}{}_{M}\gamma^{C}{}_{N}\gamma^{D}{}_{K}\,^{(D)} R^{A}{}_{BCD} = &
        \,^{(D-1)}R^{L}{}_{MNK} + K_{MK} K^{L}{}_{N} - K_{KN} K^{L}{}_{M}
\,,\\
\gamma^{A}{}_{M}\gamma^{B}{}_{N}\gamma^{C}{}_{K}\,^{(D)}R_{LABC} n^L = &
        D_{N} K_{MK} - D_M K_{KN}
\,,\\
\gamma^{A}{}_{M}\gamma^{B}{}_{N} \,^{(D)}R_{LAKB} n^L n^K = &
        \Lie_{n} K_{MN} + K_{ML} K^{L}{}_{N} + \frac{1}{\alpha} D_{M} D_{N} \alpha
\,.
\end{align}
\end{subequations}
We will use these relations in the next section to derive the time evolution form of 
Einstein's equations.

\noindent{{\bf{Decomposition of Einstein's equations:}}}
The dynamical degrees of freedom in Einstein gravity are determined by the EoMs~\eqref{eq:EEs}.
In order to study the dynamics in the high curvature regime of 
gravity, such as collisions of BHs or their stability including backreaction onto the spacetime,
we have to solve these numerically.
Typically, the EoMs are cast into a Cauchy problem, i.e., they are rewritten as a set 
of (non-linear) PDEs that are first order in time and second order in space.
In this section I will sketch the derivation of this formalism.
This section is accompanied by a \textsc{ mathematica} notebook ``GR\_Split.nb'' 
which is available online~\cite{ConfWeb}. 
The notebook makes extensive use of the freely available \textsc{xtensor} package
developed by J.~M.~Mart\'{i}n-Garc\'{i}a~\cite{xtensor}.
For convenience let us rewrite Eqs.~\eqref{eq:EEs} in the form 
$E_{MN}=0$~\footnote{This is not strictly necessary, but will make our life easier 
when keeping track of all the derived expressions.}.
Specifically, we get
\begin{subequations}
\label{eq:EEs2}
\begin{align}
\label{eq:EEs21}
E_{1,MN} = & \,^{(D)}R_{MN} - \frac{1}{2} g_{MN}\,^{(D)}R - 8\pi G_{D} T_{MN}
\,,\quad\text{or}\\
\label{eq:EEs22}
E_{2,MN} = & \,^{(D)}R_{MN} - 8 \pi G_{D} \left( T_{MN} - \frac{1}{2} g_{MN} T \right)
\,,
\end{align}
\end{subequations}
where I restrict myself to asymptotically flat spacetimes, i.e., $\Lambda = 0$.
As discussed in Sec.~\ref{ssec:DecompST}, we can decompose any tensor into
its spatial, normal and mixed components. 
Let us first consider the various projections of the energy momentum tensor $T_{MN}$.
Employing Eqs.~\eqref{eq:TensProj} yields 
\begin{align}
\label{eq:ProjTmn}
& 
T_{MN} n^M n^N =  \rho
\,,\qquad
\gamma^{K}{}_{\bm} T_{KN} n^{N} = - j_{\bm} 
\,,\qquad
\gamma^{K}{}_{\bm} \gamma^{L}{}_{\bn} T_{KL} = S_{\bm\bn}
\,,
\end{align}
where $\rho$ is the energy density, $j_{\bm}$ is the energy-momentum flux and 
$S_{\bm\bn}$ are the spatial components of the energy-momentum tensor.
Next, we perform the split prescribed by Eqs.~\eqref{eq:TensProj} for Einstein's Equations,
where we will make use of the Gauss-Codazzi equations~\eqref{eq:GaussCodazzi}.
If we contract Eq.~\eqref{eq:EEs21} twice with the normal vector, 
we obtain the \textit{Hamiltonian} constraint
\begin{align}
\label{eq:ADMHam0}
\H = & E_{1,MN} n^{M} n^{N} = 
\,^{(D-1)}R - K_{\bm\bn} K^{\bm\bn} + K^2 - 16\pi G_{D} \rho = 0
\,,
\end{align}
where $\,^{(D-1)}R$ is the $(D-1)$-dimensional Ricci scalar and $K=\gamma^{\bm\bn} K_{\bm\bn}$ is the trace
of the extrinsic curvature.
Considering the mixed projections of Eq.~\eqref{eq:EEs21} yields the \textit{momentum} constraint
\begin{align}
\label{eq:ADMmom0}
\M_{\bm} = & - \gamma^{K}{}_{\bm} E_{1,KN} n^{N} =
D_{\bn} K^{\bn}{}_{\bm} - D_{\bm} K - 8\pi G_{D} j_{\bm} = 0
\,.
\end{align}
The fully spatial projection of Eq.~\eqref{eq:EEs22} becomes
\begin{align}
\label{eq:ADMdtKmn0}
\P_{\bm\bn} = & \gamma^{K}{}_{\bm} \gamma^{L}{}_{\bn} E_{2,KL} 
= 0
\nonumber\\
= &
- \Lie_{n} K_{\bm\bn}
- \frac{1}{\alpha} D_{\bm}D_{\bn}\alpha + \,^{(D-1)}R_{\bm\bn} 
- 2 K^{\bk}{}_{\bm} K_{\bk\bn} + K^2 K_{\bm\bn} 
\nonumber \\ &
+ 4 \pi G_{D} \left( \gamma_{\bm\bn} ( S - \rho ) - 2 S_{\bm\bn} \right)
\,,
\end{align}
with $S=\gamma^{\bm\bn} S_{\bm\bn}$. The first term of the right-hand-side of the equation is the Lie derivative
of the extrinsic curvature along the normal vector $n^M$ and, 
because of  Eq.~\eqref{eq:talpbet}, involves the time derivative of $K_{\bm\bn}$,
thus providing its time evolution equation. 
The Einstein's equations in $(D-1)+1$ form are given by 
Eqs.~\eqref{eq:ADMHam0},~\eqref{eq:ADMmom0},~\eqref{eq:ADMdtKmn0} 
together with the relation~\eqref{eq:def2Kmn}. 
To summarize, let us rewrite them explicitly as time evolution equations~\footnote{ 
In the reminder of this section I will only refer to $(D-1)$-dimensional quantities and will therefore
drop the superscripts $\,^{(D-1)}X$ and $\,^{(D)}X$. Furthermore, I now set $G_D=1$.}
\begin{subequations}
\label{eq:ADMeqs}
\begin{align}
\label{eq:ADMHam}
\H = & R - K_{\bm\bn} K^{\bm\bn} + K^2 - 16 \pi \rho = 0 
\,,\\
\label{eq:ADMmom}
\M_{\bm} = & D_{\bn} K^{\bn}{}_{\bm} - D_{\bm} K - 8 \pi j_{\bm} = 0
\,,\\
\label{eq:ADMdtGam}
\p_t \gamma_{\bm\bn} = & - 2 \alpha K_{\bm\bn} + \Lie_{\beta} \gamma_{\bm\bn}
\,,\\
\label{eq:ADMdtKmn}
\p_t K_{\bm\bn} = &
- D_{\bm}D_{\bn}\alpha 
+ \alpha\left( R_{\bm\bn} - 2 K^{\bk}{}_{\bm} K_{\bk\bn} + K K_{\bm\bn} \right)
\nonumber \\ &
+ 8 \pi \alpha \left( \frac{1}{D-2}\gamma_{\bm\bn} ( S - \rho ) - S_{\bm\bn} \right)
+ \Lie_{\beta} K_{\bm\bn}
\,.
\end{align}
\end{subequations}
The NR community often dubs Eqs.~\eqref{eq:ADMeqs}  \textit{ADM equations}, refering to 
Arnowitt, Deser and Misner~\cite{Arnowitt:1962hi}, 
although their original work used the Hamiltonian formalism
and the equations in the above form have been derived by York~\cite{York:1979} in $D=4$.

The first two equations~\eqref{eq:ADMHam} and~\eqref{eq:ADMmom} are the physical constraints of the system.
They consist of $D$ coupled elliptic PDEs which are in general hard to solve.
Therefore, in so-called free evolution schemes, the constraints are solved only on the initial timeslice
and monitored throughout the evolution as consistency check of a simulation.
For details of various techniques to solve the constraints 
and approaches to construct initial data for the spatial metric and extrinsic curvature 
I refer the interested reader to Cook's article~\cite{Cook:2000vr}
(in $4$-dimensional asymptotically flat spacetimes)
and Okawa's contribution to these lecture notes~\cite{Okawa:2013afa} for $D$-dimensional spacetimes.

The second set of equations~\eqref{eq:ADMdtGam} and~\eqref{eq:ADMdtKmn}
represent the time evolution equations for $(\gamma_{\bm\bn},K_{\bm\bn})$ and encode the dynamics of the system. 
In the following I will engage in a further discussion of these PDEs.

\noindent{{\bf{BSSN formulation of Einstein's equations:}}}
Although the ADM-York formalism of Einstein's equations as a time evolution problem, Eqs.~\eqref{eq:ADMeqs}, 
has been around since the late 1970's~\cite{York:1979},
the 2-body problem in GR has only been solved in 2005 by Pretorius~\cite{Pretorius:2005gq,Pretorius:2006tp} 
followed by Baker et al.~\cite{Baker:2005vv} 
and Campanelli et al.~\cite{Campanelli:2005dd} in 2006.
Pretorius' seminal work~\cite{Pretorius:2005gq,Pretorius:2006tp} has been based on a 
Generalized Harmonic formulation in which the EoMs are basically written 
as a set of wave equations for the metric. 
Shortly afterwards, Baker et al.~\cite{Baker:2005vv} and Campanelli et al.~\cite{Campanelli:2005dd}
independently presented 
successful numerical simulations of orbiting and merging BH binaries 
where they have used a modified version of Eqs.~\eqref{eq:ADMeqs} 
as introduced by Baumgarte \& Shapiro~\cite{Baumgarte:1998te} 
and Shibata \& Nakamura~\cite{Shibata:1995we} \textit{(BSSN)}
together with a particular choice
for the gauge variables and treatment of the BH singularity nowadays known as 
\textit{moving puncture approach}~\cite{Brandt:1997tf,Alcubierre:2002kk,Baker:2005vv,Campanelli:2005dd}.
Retrospectively, it has been this particular combination of ``ingredients'' 
-- although already known on their own and inspired, e.g., by neutron star simulations -- which 
led to the breakthrough~\footnote{Very much like flour, eggs, butter and sugar 
on their own don't make a delicious cake, but properly combining and baking them does.}.
Here, I will focus on the latter method which is commonly referred to as \textit{BSSN method with moving punctures}
and recommend 
Refs.~\cite{Pretorius:2005gq,Pretorius:2006tp,Pretorius:2004jg,Lindblom:2005qh,Szilagyi:2009qz,
Lehner:2011wc}
to learn more about the GHG formalism.

In hindsight it is not surprising that the ADM formalism failed:
From a mathematical perspective, one can show that the underlying PDE system is only 
weakly hyperbolic~\cite{Alcubierre:2008, Sarbach:2012pr},
which means that it is an ill-posed initial value problem 
and therefore prone to numerical instabilities.
As Hilditch discusses in great detail in his contribution to these lecture notes~\cite{Hilditch:2013sba},
a strongly hyperbolic initial value formulation of the PDE system
is a necessary condition to obtain a stable numerical scheme.
By now, there is a plethora of well-posed initial value formulations of Einstein's equations.
The most commonly used 
version is the BSSN formulation~\cite{Shibata:1995we,Baumgarte:1998te}
or variations thereof with, e.g., 
modified dynamical variables~\cite{Witek:2010es,Witek:2013koa},
additional constraint damping schemes called 
\textit{Z4c}~\cite{Bernuzzi:2009ex, Weyhausen:2011cg, Cao:2011fu, Hilditch:2012fp, Alic:2011gg}
or covariant formulations~\cite{Brown:2009dd,Brown:2009ki}.
In the following I will summarize the main aspects of the (original) BSSN formulation.

The key is to change the character of the PDEs~\eqref{eq:ADMdtGam} and~\eqref{eq:ADMdtKmn} 
such that they become
a well-posed initial value formulation~\cite{Alcubierre:2008, 
Gustafsson1995, Sarbach:2012pr, Gundlach:2006tw, Hilditch:2010wp, Hilditch:2013ila}.
We will accomplish this goal by adding the constraints~\eqref{eq:ADMHam} and~\eqref{eq:ADMmom}
to the evolution equations
and performing a conformal decomposition of the dynamical variables $(\gamma_{\bm\bn},K_{\bm\bn})$.
The new set of dynamical variables are 
the conformal factor $\chi$~\footnote{Alternatively, also the conformal
factors $\phi=-\tfrac{1}{4}\ln\chi=\tfrac{1}{4(D-1)}\ln\gamma$ or
$W=\sqrt{\chi} = \gamma^{-\tfrac{1}{2(D-1)}}$ are used.} 
and metric $\tg_{\bm\bn}$,
the trace $K$ of the extrinsic curvature and its conformally decomposed trace-free part $\tA_{\bm\bn}$
and the conformal connection function $\tG^{\bm}$
\begin{subequations}
\label{eq:BSSNvars}
\begin{align}
\label{eq:BSSNmetric}
\chi = & \gamma^{-\frac{1}{D-1}}
\,,\qquad 
\tg_{\bm\bn} = \gamma^{-\frac{1}{D-1}} \gamma_{\bm\bn} = \chi \gamma_{\bm\bn}
\,,\\
\label{eq:BSSNKA}
K = & \gamma^{\bm\bn} K_{\bm\bn}
\,,\quad
\tA_{\bm\bn} = \chi A_{\bm\bn} = \chi\left( K_{\bm\bn} - \frac{1}{D-1} \gamma_{\bm\bn}K \right)
\,,\\
\label{eq:BSSNGam}
\tG^{\bm} = & \tg^{\bk\bl} \tG^{\bm}{}_{\bk\bl} = - \p_{\bk} \tg^{\bm\bk}
\,,
\end{align}
\end{subequations}
with $\gamma=\det\gamma_{\bm\bn}$ and $\tg = \det\tg_{\bm\bn} =1$. 
The conformal connection function $\tG^{\bm}$ has been introduced such that the Ricci scalar can 
be rewritten as a Laplace operator for the conformal metric while its remaining derivatives are absorbed
into the new variable.
Be aware, that the BSSN variables $(\chi, \tg_{\bm\bn}, \tA_{\bm\bn})$  are tensor 
densities $\tilde{T}=\gamma^{\frac{\mathcal{W}}{2}} T$ of weight
$\mathcal{W}=-\tfrac{2}{D-1}$, while $\tG^{\bm}$ is, in fact, the derivative of a tensor density
and transforms as 
\begin{align}
\label{eq:densGam}
\tG^{\bm} = & -\gamma^{\frac{\mathcal{W}}{2}} \left( \mathcal{W} \gamma^{\bm\bn}\Gamma^{\bk}{}_{\bk\bn}+\p_{\bn}\gamma^{\bm\bn}\right)
\,.
\end{align}
Note, that the transformations~\eqref{eq:BSSNvars}  
add auxiliary algebraic and differential constraints to the system
\begin{subequations}
\begin{align}
\label{eq:BSSNauxconst}
\D = & \ln(\tg) = 0 
\,,\qquad
\T = \tg^{\bm\bn} \tA_{\bm\bn} = 0
\,,\\
\label{eq:BSSNGconstr}
G_{\bm} = & \tg_{\bm\bn} \tG^{\bn} - \tG^{\rm{[DEF]}}_{\bm} = \tg_{\bm\bn} \tG^{\bn} - \tg^{\bk\bl} \p_{\bk}\tg_{\bm\bl} = 0
\,.
\end{align}
\end{subequations}
While traditionally in NR textbooks (see, e.g., Alcubierre~\cite{Alcubierre:2008}) 
the BSSN formulation is now derived by inserting the 
transformations~\eqref{eq:BSSNvars} into the ADM equations~\eqref{eq:ADMeqs} and substituting 
the divergence of $\tA_{\bm\bn}$ and the Ricci scalar with the constraints
I here take a different route. 
Following Ref.~\cite{Witek:2013koa} 
I kick off by first adding the constraints~\eqref{eq:ADMHam},~\eqref{eq:ADMmom} and~\eqref{eq:BSSNGconstr}
to the ADM evolution equations~\eqref{eq:ADMdtGam} and~\eqref{eq:ADMdtKmn}. 
Thus, the modified structure of the evolution PDE system becomes immidiately evident
and writes
\begin{subequations}
\label{eq:BSSNevol0}
\begin{align}
\label{eq:BSSNdtg0}
\p_t\gamma_{\bm\bn} = & \rm{[ADM]}
\,,\\
\label{eq:BSSNdtKmn0}
\p_t K_{\bm\bn} = & {\rm{[ADM]}} + \alpha \p_{(\bm} G_{\bn)} 
                 - \frac{1}{D-1} \alpha \gamma_{\bm\bn} \left( \H + \gamma^{\bk\bl}\p_{\bk}G_{\bl} \right)
\,,\\
\label{eq:BSSNdtG0}
\p_t G_{\bm} = & 2\alpha\M_{\bm} -2\alpha G^{\bn} A_{\bm\bn} + \Lie_{\beta} G_{\bm} + \gamma_{\bm\bn} G^{\bk} \p_{\bk}\beta^{\bn}
                - \frac{2}{D-1} G_{\bm} \p_{\bn}\beta^{\bn}
\,,
\end{align}
\end{subequations}
where ``[ADM]'' denotes the ADM equations~\eqref{eq:ADMeqs}.
I progress by inserting the new dynamical variables, Eqs.~\eqref{eq:BSSNvars},
into our modified evolution equations~\eqref{eq:BSSNevol0}. This procedure yields
the BSSN equations
\begin{subequations}
\label{eq:BSSN}
\begin{align}
\label{eq:BSSNdtchi}
\p_t \chi = & \frac{2}{D-1} \alpha \chi K + \beta^{\bm}\p_{\bm}\chi -\frac{2}{D-1}\chi \p_{\bm} \beta^{\bm}
\,,\\
\label{eq:BSSNdtg}
\p_t\tg_{\bm\bn} = & -2\alpha \tA_{\bm\bn} + 2 \p_{(\bm}\beta_{\bn)} - \frac{2}{D-1}\tg_{\bm\bn} \p_{\bk}\beta^{\bk}
\,,\\
\label{eq:BSSNdtK}
\p_t K = & - D^{\bk} D_{\bk} \alpha + \alpha \left( \tA_{\bk\bl} \tA^{\bk\bl} + \frac{K^2}{D-1}\right) + \beta^{\bk} \p_{\bk} K 
        + \frac{8\pi \alpha}{D-2} \left( (D-3) \rho + S \right)
\,,\\
\label{eq:BSSNdtA}
\p_t\tA_{\bm\bn} = & -\chi [D_{\bm\bn}\alpha]^{\rm{tf}} + \alpha \chi [R_{\bm\bn}]^{\rm{tf}} +\alpha(K\tA_{\bm\bn} -2\tA^{\bk}{}_{\bm} \tA_{\bk\bn})
        + 8\pi\alpha\chi [S_{\bm\bn}]^{\rm{tf}}
\nonumber\\ &
        + 2\tA_{\bk(\bm} \p_{\bn)} \beta^{\bk} + \beta^{\bk}\p_{\bk}\tA_{\bm\bn} -\frac{2}{D-1} \tA_{\bm\bn} \p_{\bk}\beta^{\bk}
\,,\\
\label{eq:BSSNdtG}
\p_t\tG^{\bm}=&
- 2 \frac{D-2}{D-1} \alpha \tg^{\bm\bk} \tD_{\bk}K
- 2 \tA^{\bm\bk} \tD_{\bk}\alpha
-   \alpha \frac{D-1}{\chi} \tA^{\bm\bk} \tD_{\bk}\chi 
+  2 \alpha \tG^{\bm}{}_{\bk\bl} \tA^{\bk\bl}
\nonumber \\ &
+ \beta^{\bk} \p_{\bk} \tG^{\bm}
- \tG^{\bk} \p_{\bk} \beta^{\bm}
+ \frac{2}{D-1} \tG^{\bm} \p_{\bk}\beta^{\bk} 
+ \frac{D-3}{D-1} \tg^{\bm\bk} \p_{\bk}\p_{\bl} \beta^{\bl} 
+ \tg^{\bk\bl} \p_{\bk}\p_{\bl} \beta^{\bm}
\nonumber \\ &
- 16 \pi \frac{\alpha}{\chi} j^{\bm}
\,,
\end{align}
\end{subequations}
where $[X_{\bm\bn}]^{\rm{tf}} = X_{\bm\bn} - \frac{1}{D-1}\gamma_{\bm\bn}X$
denotes the trace-free part with respect to the \textit{physical} metric $\gamma_{\bm\bn}$.
$D_{\bm}$ and $\tD_{\bm}$ are the covariant derivative associated with,
respectively, the physical metric $\gamma_{\bm\bn}$ and conformal metric $\tg_{\bm\bn}$.
The respective Levi-Civita connections are related via
\begin{align}
\Gamma^{\bm}{}_{\bk\bl} = &  \tG^{\bm}{}_{\bk\bl} 
                           - \frac{1}{2\chi}\left(\tg^{\bm}{}_{\bk}\p_{\bl}\chi + \tg^{\bm}{}_{\bl}\p_{\bk}\chi 
                                - \tg_{\bk\bl} \tg^{\bm\bn} \p_{\bn}\chi \right)
\end{align}
The Ricci tensor $R_{\bm\bn}$ transforms as
\begin{align}
R_{\bm\bn} = & \tR_{\bm\bn} + \frac{D-3}{2\chi}\tD_{\bm}\tD_{\bn}\chi +\frac{1}{2\chi}\tg_{\bm\bn}\tg^{\bk\bl} \tD_{\bk}\tD_{\bl}\chi
\nonumber\\ &
        -\frac{D-3}{4\chi^2}\tD_{\bm}\chi\tD_{\bn}\chi -\frac{D-1}{4\chi^2}\tg_{\bm\bn}\tg^{\bk\bl}\tD_{\bk}\chi\tD_{\bl}\chi
\,,
\end{align}
with $\tR_{\bm\bn}$ being the Ricci tensor with respect to the conformal metric.
The second derivative of the lapse function writes
\begin{align}
D_{\bm} D_{\bn}\alpha = & \tD_{\bm}\tD_{\bn} \alpha 
                         + \frac{1}{ \chi} \tD_{(\bm}\chi \tD_{\bn)} \alpha 
                         - \frac{1}{2\chi} \tg_{\bm\bn}\tg^{\bk\bl} \tD_{\bk}\chi\tD_{\bl} \alpha
\,.
\end{align}
In order to close the PDE system~\eqref{eq:BSSN} we have to specify 
the coordinate gauge functions $(\alpha,\beta^{\bm})$.
While GR, in principle, admits coordinate degrees of freedom, 
the particular specification has a tremendous impact on the stability of 
a numerical simulation.
For example, if we were to use the most obvious choice with
$(\alpha=1,\beta^{\bm}=0)$, known as geodesic slicing,
any geodesic would reach the BH singularity in finite time
thus yielding any NR simulation to break down and terminate.
Succesful simulations of BH spacetimes have combined the BSSN equations~\eqref{eq:BSSN}
with the so-called \textit{moving puncture} gauge~\cite{Brandt:1997tf,Baker:2005vv,Campanelli:2005dd,Alcubierre:2008}
employing the \textit{1+log slicing} condition for the lapse function and the \textit{$\Gamma$-driver}
condition for the shift vector. I here restrict myself to the version presented in Ref.~\cite{vanMeter:2006vi}.
The puncture gauge generalizes to higher dimensions as~\cite{Yoshino:2009xp,Shibata:2009ad,Shibata:2010wz}
\begin{subequations}
\label{eq:MovingPuncture}
\begin{align}
\label{eq:1log}
\p_t \alpha = & -2\eta_{\alpha} \alpha K + \beta^{\bk} \p_{\bk}\alpha
\,,\\
\label{eq:GammaDriver}
\p_t \beta^{\bm} = & \frac{D-1}{2(D-2)}\zeta_{\Gamma} \tG^{\bm} - \eta_{\beta} \beta^{\bm} + \beta^{\bk}\p_{\bk} \beta^{\bm}
\,,
\end{align}
\end{subequations}
where $\eta_{\beta}\beta^{\bm}$ is a damping term and $\beta^{\bk}\p_{\bk}X$ are the advection terms.
This choice of the lapse causes the slices to evolve slower close to the BH singularity
and thus avoids ``touching'' it. This \textit{singularity avoiding} property allows 
for long-term stable BH evolutions.
The $\Gamma$-driver shift condition is a generalization of the original puncture gauge~\cite{Brandt:1997tf,vanMeter:2006vi}
and allows the coordinates to adapt to the movement of the BHs over the numerical domain.
%

\newpage
\subsection{Exercises}
\subsubsection{Properties of the extrinsic curvature}
\label{ssec:ExcPropCurv}
Show that the extrinsic curvature, defined in Eq.~\eqref{eq:def1Kmn},
is 
\begin{enumerate}
\item purely spatial, i.e., $K_{MN} n^M = 0$.
\item related to the spatial metric $\gamma_{\bm\bn}$ by Eq.~\eqref{eq:def2Kmn}.
Make use of the following definition of the Lie derivative of a tensor $T^{M_1...M_p}{}_{N_1...N_q}$
along a vector $u^L$
\begin{align}
\label{eq:DefLie}
\Lie_{u} T^{M_1...M_p}{}_{N_1...N_q} = & u^{L} \nabla_L T^{M_1...M_p}{}_{N_1...N_q}
\nonumber \\ &
        - T^{L...M_p}{}_{N_1...N_q} \nabla_L u^{M_1} - ...
\nonumber \\ &
        + T^{M_1...M_p}{}_{L...N_q} \nabla_{N_1} u^L + ...
\,;
\end{align}
\end{enumerate}

\subsubsection{$(D-1)+1$ form of Einstein's equations with cosmological constant}
\label{ssec:EELam}
Perform the ADM decomposition of Einstein's equations~\eqref{eq:EEs} 
including a cosmological constant.
For this purpose you can modify the \textsc{mathematica} notebook
``GR\_Split.nb'' available on the conference webpage~\cite{ConfWeb}.
\subsubsection{$(D-1)+1$ form of Einstein's equations in a non-vacuum spacetime}
\label{ssec:EESca}
Derive the constraint and time evolution equations for Einstein's equations 
with a non-vanishing energy momentum tensor.
Exemplarily, let us consider the energy momentum tensor for a real scalar field $\Phi$ which is minimally coupled to GR
and given by
\begin{align}
T_{MN} = & \nabla_M \Phi \nabla_N \Phi - \frac{1}{2} g_{MN} \nabla^{L}\Phi \nabla_{L}\Phi
\,.
\end{align}
Note, that the system will be closed by a EoM for the scalar field which is given by 
the energy-momentum conservation
\begin{align}
\nabla_{M} T^{MN} = & 0
\,.
\end{align}
For this purpose you can modify the \textsc{mathematica} notebook ``GR\_Split.nb'' available online~\cite{ConfWeb}.

\newpage
\section{Interludium}

In the previous section I have derived the general $(D-1)+1$-dimensional BSSN formulation of
Einstein's equations, which reduces to the well-know expressions for $D=4$~\cite{Alcubierre:2008}.
At first glance the necessary ingredients appear to be in place for a 
straightforward implementation using, e.g., the \textit{method of lines} 
(MoL)~\footnote{The MoL is a technique, in which first all spatial quantities are evaluated on
a timeslice and then evolved in time using a standard (time) integrator such as the
$4^{\rm{th}}$ order Runge-Kutta scheme.}.
However, celebrations would be premature. 
Bear in mind that the computational requirements increase with dimensionality. 
Evolving a $D$-dimensional spacetime results in an increasingly large number of
grid functions: If we count only the BSSN variables on one timeslice they result in
$2D+1+2\sum_{i=1}^{D-1}i$ grid functions~\footnote{To give an example: 
in $D=4$ the BSSN variables result in $29$ grid functions, while the number 
increases to, respectively, $41$ and $55$ in $D=5$ and $D=6$.}.
For the time evolution using, e.g., the $4^{\rm{th}}$ order Runge-Kutta time integrator
we need to store these functions on $3$ time levels. Additionally, there are 
the ADM variables and a vast number
of auxiliary grid functions to be stored.
Furthermore, in the naive approach all these functions would have to be evaluated 
on grids of size $N^{D-1}$, with $N$ being the number of points in one direction.
In order to reduce these computational requirements such that they are 
feasible for up-to-date computational resources,
we need to reduce our EoMs to \textit{at most}
$3+1$-dimensional problems.
This implies considering scenarios with a $SO(D-2)$ or $SO(D-3)$ isometry.
While this simplification does constrain the phase-space of possible scenarios, 
it still allows us to investigate many interesting higher dimensional phenomena, such as: 
head-on collisions of BHs with varying initial boost
and impact parameter~\cite{Zilhao:2010sr,Witek:2010xi,Witek:2010az,Zilhao:2011yc,Okawa:2011fv},
Myers-Perry BHs with one~\cite{Yoshino:2009xp,Shibata:2009ad,Shibata:2010wz} or more spin parameters,
and black strings as well as
dynamical instabilities~\cite{Choptuik:2003qd,Lehner:2010pn,Lehner:2011wc},
such as the Gregory-Laflamme instability~\cite{Gregory:1993vy,Gregory:2011kh}.

A further advantage of formulating our higher dimensional task 
as an effectively $2+1$- or $3+1$- problem for any $D$ 
is that we can develop a code capable of dealing 
with generic spacetime dimension, where $D$ is just a parameter,
instead of implementing a new version for every change in dimensionality.

A straightforward approach would be a direct implementation using the considered symmetries.
While \textit{in principle} this can be done, we would always have to deal with 
coordinate singularities at the origin or axis of symmetry which are sometimes 
difficult to treat~\cite{Rinne:2005df}.
Therefore, the NR community embraces Cartesian coordinates which are simpler to handle and avoid
the aforementioned coordinate singularities.
Naturally, we now have to provide some smart way to combine both: 
numerically very robust Cartesian coordinates and 
the necessary symmetries of our particular tasks.

The literature on NR in higher dimensional BH spacetimes knows two very successful approaches:
\noindent{(i)} the so-called \textit{Cartoon} method~\cite{Okawa:2011fv, 
Yoshino:2009xp, Shibata:2010wz, Lehner:2010pn, Lehner:2011wc}
and \noindent{(ii)} a formulation based on the \textit{dimensional reduction} 
by isometry~\cite{Zilhao:2010sr, Witek:2010xi}.
In the following sections, I will discuss the key aspects of both approaches.

\section{The Cartoon method}\label{sec:Cartoon}
The \textit{Cartoon} method, short for ``\textit{Cart}esian 
\textit{two}dimensional''~\footnote{The name also was inspired by 
the typical (low-budget) TV cartoons that animate the $3d$ world
in $2\tfrac{1}{2}$ dimensions.}
was originally developed to investigate head-on collisions of BHs 
in $D=4$~\cite{Alcubierre:1999ab}.
The natural choice of coordinates to study these axissymmetric configurations
would be polar coordinates $(\rho,\varphi,z)$
allowing to evolve the system on a grid with $N^2$ points.
However, this choice exhibits a coordinate singularity at the axis of symmetry
which is sometimes difficult to treat and might cause numerical instabilities.
Since this coordinate singularity is absent in Cartesian coordinates,
they often are the preferred choice of grid coordinates in heavy NR simulations.
On the other hand, because the symmetry of the setup does not obviously emerge in Cartesian 
coordinates, they would require to evolve a grid made up of $N^{D-1}$ points thus demanding more 
computational resources.
Now, the idea behind the Cartoon method is to combine the advantages of both choices
to reduce the numerical costs.
%
\subsection{The Cartoon method in $D=4$ revisited}\label{ssec:Cartoon4D}
\begin{figure}[htpb!]
\begin{center}
\includegraphics[width=0.75\textwidth]{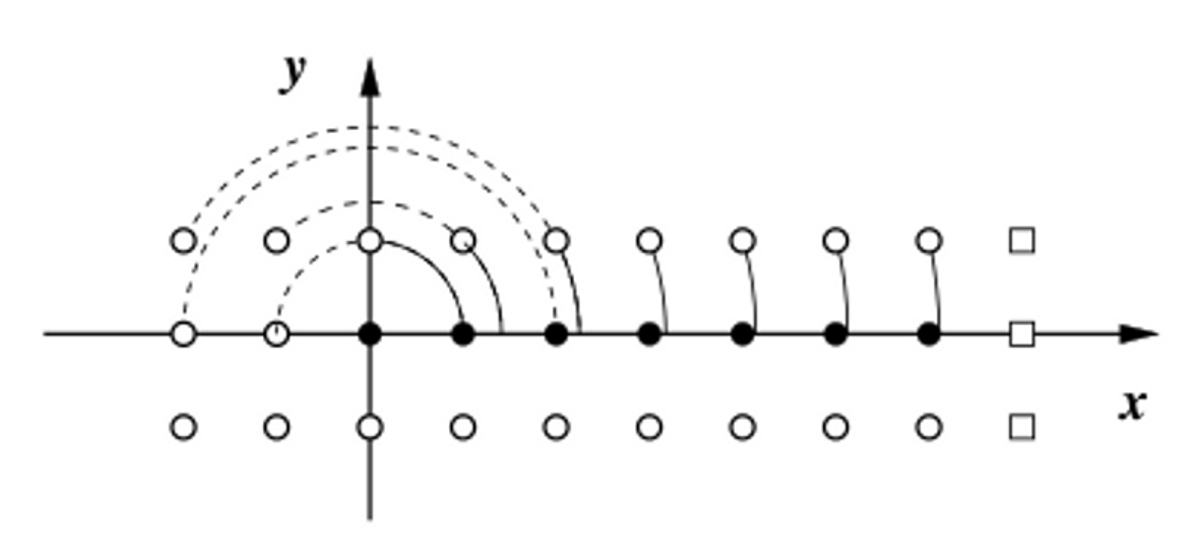}
\end{center}
\vspace*{8pt}
\caption{
\label{fig:Cartoon4D}
Illustration of the Cartoon method on a typical numerical grid around $y=0$. 
Note, that the $z$-direction is supressed for visibility.
The interior grid $0\leq x \le x_{\rm{max}}$ is marked by solid points, on which we employ the standard,
centered FD stencils.
At the outer points (marked by squares) we impose the physical boundary conditions.
The arcs visualize the rotation allowed by axissymmetry to obtain the grid functions at points $y\neq0$
and $x\le0$ (open circles).
Taken from Alcubierre et al.~\cite{Alcubierre:1999ab}.
}
\end{figure}

To elaborate the main aspects of the Cartoon method let us first consider 
a BH head-on collision along the $z$-axis in $D=4$ dimensions~\cite{Alcubierre:1999ab}.
Then, the system has a $U(1)$ symmetry around the $z$-axis with Killing vector (KV) field
$\p_{\varphi}=x\p_{y}-y\p_{x}$
and the dynamics of the configuration are confined to the $x-z$-plane, i.e., $y=0$.
However, because the symmetry of the problem is not explicit in Cartesian coordinates, 
we still have to take derivatives in all spatial directions.
In general, the derivative of a function $u$ with respect to $y$ does not vanish.
Numerically, we evaluate this derivative by employing (centered) finite difference (FD) stencils
which are given by~\cite{1992nrca,Gustafsson1995,Alcubierre:2008}
\begin{subequations}
\label{eq:FDstencils}
\begin{align}
\label{eq:FD2nd}
\p_y u_{j} = & \frac{1}{2\Delta y}\left( u_{j+1} - u_{j-1} \right)
\qquad\qquad\qquad\qquad\qquad 
(2^{\rm{nd}}\text{order stencil)}
\,,\\
\label{eq:FD4th}
\p_y u_{j} = & \frac{1}{12\Delta y}\left( u_{j-2} - 8 u_{j-1} + 8 u_{j+1} - u_{j+2} \right)
\qquad (4^{\rm{th}}\text{order stencil)}
\,,
\end{align}
\end{subequations}
where $\Delta y$ denotes the grid spacing in $y$.
Inspecting Eqs.~\eqref{eq:FDstencils} we observe that we do not have to evolve the full
$3$-dimensional domain, but only $(2g+1)$ grid-points in the $\pm y$ direction in the neighborhood of $y=0$.
Here, $g$ depends on the order of the FD scheme and is, respectively, 
$g=1$ for second and $g=2$ for fourth order FD stencils.
This observation allows us to reduce the grid size from a cube with $N^3$ grid points
to a slab or cuboid with $(2g+1) N^2$ grid points.
Now, we evaluate the functions at grid points $(x,y=0,z)$ using
centered FD stencils in the interior region
and employ physical boundary conditions at the outer points,
as is illustrated in Fig.~\ref{fig:Cartoon4D}.
Additionally, we need to populate the points at $y\neq0$.

The strategy that we use consists of the following steps~\cite{Alcubierre:1999ab}
\begin{enumerate}
\item evaluation of grid functions at points $(x,y=0,z)$
\item interpolation of function values at grid points to points\\ 
      $p=(\rho,\varphi=0,z)=(\sqrt{x^2+y^2},0,z)$
\item rotation of tensors from a point $p$ to a grid point $p'=(x,y\neq0,z)$.
\end{enumerate}
We discuss each of these items in more detail below.

\noindent{{\bf{Rotation of tensors:}}}
Let us recall that for now we consider a $3$-dimensional space $\,^{(3)}\Sigma_t$
which exhibits a $U(1)$ symmetry around the z-axis 
and choose the plane for which $y=0$.
The natural coordinates for this kind of problem are polar coordinates $(\rho,\varphi,z)$
which are related to Cartesian coordinates by 
\begin{align}
\label{eq:CartPol}
x = \rho \cos\varphi\,,\quad 
y = \rho \sin\varphi\,,\quad
z = z\,,\quad\text{with}\quad
\rho = \sqrt{x^2+y^2}
\,.
\end{align}
Let us consider the rotation of a spatial tensor field $T \in\,^{(3)}\Sigma_t$ 
by an angle $-\varphi_{0}$ around the z-axis.
This is equivalent to keeping the tensor field fixed and instead rotate the coordinates by an
angle $+\varphi_{0}$.
The rotation defines a diffeomorphism 
$R:~\,^{(3)}\Sigma_t \rightarrow \,^{(3)}\Sigma_t$ with
$R: T \rightarrow R^{\ast} T$
and the rotation matrix is given by 
\begin{align}
\left( R(\varphi_0)^{i}{}_{j} \right) = &
\left( \begin{matrix}
\cos\varphi_0 & -\sin\varphi_0 & 0 \\ \sin\varphi_0 & \cos\varphi_0 & 0 \\ 0 & 0 & 1 
\end{matrix} \right)
= \left( \begin{matrix}
\frac{x}{\rho} & -\frac{y}{\rho} & 0 \\ \frac{y}{\rho} & \frac{x}{\rho} & 0 \\ 0 & 0 & 1 
\end{matrix} \right)
\end{align}
Note, that $R(\varphi_0)^{-1}=R(-\varphi_0)$.
The transformation of any $(p,q)$-tensor $T$ 
with $T(p) \rightarrow R^{\ast}T(R(p))$
from a point $p=(\rho,0,z)$ to a point $p'=R(p)=(x,y,z)$
is described by
\begin{align}
\label{eq:TensorTrafo4D}
T^{i_1...i_p}{}_{j_1...j_q}(x,y,z) = & R^{i_1}{}_{k_1}...R^{i_p}{}_{k_p} (R^{-1})^{l_1}{}_{j_1}...(R^{-1})^{l_q}{}_{j_q}
                                       T^{k_1...k_p}{}_{l_1...l_q} (\rho,0,z)
\,,
\end{align}
where we have already imposed the symmetry, i.e., $R^{\ast}T = T$.
For concreteness, let us consider some explicit examples.
In view of the BSSN equations~\eqref{eq:BSSN}, I focus on scalar-, vector- 
and 2-tensor-type variables~\footnote{Note that some of the BSSN variables are tensor \textit{densities}
or derivatives thereof
and, strictly speaking, these relations hold only for tensors. However, it is straightforward to first apply the 
tensor rotation to the ADM variables and then perform the conformal decomposition.}:
\begin{itemize}
\item a scalar $\Psi$ 
transforms as:
\begin{align}
\Psi(x,y,z) = & \Psi(\rho,0,z)
\end{align}
\item a vector $V^i$ 
transforms as:
\begin{align}
V^i(x,y,z) = & R^{i}{}_{j} V^{j}(\rho,0,z)
\end{align}
with
\begin{subequations}
\begin{align}
V^{x}(x,y,z) = & \frac{x}{\rho} V^{x}(\rho,0,z) - \frac{y}{\rho} V^{y}(\rho,0,z)
\,,\\
V^{y}(x,y,z) = & \frac{y}{\rho} V^{x}(\rho,0,z) + \frac{x}{\rho} V^{y}(\rho,0,z)
\,,\\
V^{z}(x,y,z) = & V^{z}(\rho,0,z) 
\,,
\end{align}
\end{subequations}
\item a 2-tensor $S_{ij}$ 
transforms as:
\begin{align}
S_{ij}(x,y,z) = & (R^{-1})^{k}{}_{i}(R^{-1})^{l}{}_{j} S_{kl}(\rho,0,z)
\end{align}
For a symmetric $2$-tensor we obtain explicitely 
\begin{subequations}
\begin{align}
S_{xx}(x,y,z) = & \frac{x^2}{\rho^2} S_{xx}(\rho,0,z) - 2\frac{xy}{\rho^2} S_{xy}(\rho,0,z) + \frac{y^2}{\rho^2} S_{yy} (\rho,0,z)
\,,\\
S_{xy}(x,y,z) = & \frac{xy}{\rho^2} \left(S_{xx}(\rho,0,z) - S_{yy}(\rho,0,z) \right) 
                 +\frac{x^2-y^2}{\rho^2} S_{xy}(\rho,0,z)
\,,\\
S_{xz}(x,y,z) = & \frac{x}{\rho} S_{xz}(\rho,0,z) - \frac{y}{\rho} S_{yz}(\rho,0,z)
\,,\\
S_{yy}(x,y,z) = & \frac{y^2}{\rho^2} S_{xx}(\rho,0,z) + 2\frac{xy}{\rho^2}S_{xy}(\rho,0,z)+\frac{x^2}{\rho^2} S_{yy}(\rho,0,z)
\,,\\
S_{yz}(x,y,z) = & \frac{y}{\rho}S_{xz}(\rho,0,z) + \frac{x}{\rho}S_{yz}(\rho,0,z)
\,,\\
S_{zz}(x,y,z) = & S_{zz}(\rho,0,z)
\,.
\end{align}
\end{subequations}
\end{itemize}

\noindent{{\bf{Interpolation:}}}
The transformation rule, Eq.~\eqref{eq:TensorTrafo4D}, implies that we require 
tensor values at $p=(\rho,\varphi=0,z)=(\sqrt{x^2+y^2},0,z)$ in order to
rotate the function to point $p'=(x,y\neq0,z)$.
However, it might happen that $p$ is \textit{not} a grid point as can be seen 
in Fig.~\ref{fig:Cartoon4D}.
Therefore, it is necessary to perform a $1$-dimensional interpolation to 
provide all tensor values along $\varphi=0$.
Typically, a Lagrange polynomial interpolation is employed
and it has been found that polynomials of degree $2$-$5$ yield good 
numerical results~\cite{1992nrca,Alcubierre:1999ab}.

\subsection{Cartoon method in $D=5$}\label{ssec:Cartoon5D}
The Cartoon method extends in a straightforward manner 
to $D=5$ spacetime dimensions. This will allow us to reduce $4+1$-dimensional problems
to effectively $3+1$- or $2+1$-dimensional ones, depending on the specific configuration. 
From our discussion in the previous section we can deduce a generic strategy or ``recipe'':
\begin{enumerate}
\item The first step consists in setting up our project, specifically: 
  \begin{enumerate}
  \item identify the symmetries that our problem exhibits,
  \item fix the axis/ hyperplanes of symmetry, 
  \item set up Cartesian coordinates accordingly and 
  \item identify their relation to curvi-linear coordinates;
  \end{enumerate}
\item Afterwards, we need to develop the linear map $R: \,^{(D-1)}\Sigma_t \rightarrow \,^{(D-1)}\Sigma_t$ 
      with $R: T(p) \rightarrow R^{\ast}T(R(p))$ being the rotation matrix providing the transformation 
      rules for tensors $T$; 
\item Next, we prepare the numerical data at the grid points on the axis or hyperplanes of symmetry;
\item We have to interpolate the grid functions to the corresponding points\\ 
      $p\in\,^{(D-1)}\Sigma_t$ in curvi-linear coordinates; 
\item Finally, we generate the function and tensor values at the points\\ 
      $p'=R(p)\in\,^{(D-1)}\Sigma_t$
      through tensor rotation, using Eq.~\eqref{eq:TensorTrafo4D}.
\end{enumerate}
To illustrate this strategy, I will discuss the example of  
a $5$-dimensional BH spacetime with a $U(1)$ symmetry,
modelling, e.g., BH collisions with an impact parameter~\cite{Okawa:2011fv,Yoshino:2009xp}.
Additionally, I give the example of a $5$-dimensional spacetime with 
a $U(1)\times U(1)$ symmetry, representing, e.g., Myers-Perry BH with two
spin parameters, as exercise in Sec.~\ref{ssec:ExerciseCartoon}.
Further examples have been discussed in the original 
publications~\cite{Yoshino:2009xp,Shibata:2009ad,Lehner:2010pn}
and I encourage the interested reader to follow them.

\noindent{{\bf{Example: $D=5$ dimensional spacetime with a $U(1)$ symmetry:}}}
This class of spacetimes allows us to model, e.g., BH collisions with an impact parameter in $D=5$
as an effectively $3+1$-dimensional problem.
The numerical domain is then reduced from $N^4$ grid points to $(2g+1)N^3$ which is
much more feasible in terms of numerical costs. 
For this purpose, we develop a numerical scheme using the Cartoon method
according to our ``recipe'':
\begin{enumerate}
\item As noted in the text, we intend to investigate a $5$-dimensional BH spacetime that
exhibits a $U(1)$ symmetry.
Therefore we consider Cartesian coordinates $(x,y,z,w)$ 
where we choose the $z-w$-plane as the plane of symmetry. Then, $\p_{\psi}=z\p_{w}-w\p_{z}$ is 
a KV field of the spacetime. The Cartesian coordinates are related to 
polar-like coordinates $(x,y,\rho,\psi$) via
\begin{align}
x=x\,,\quad
y=y\,,\quad
z=\rho\cos\psi\,,\quad
w=\rho\sin\psi\,,\quad\text{with}\quad
\rho = \sqrt{z^2+w^2}
\,;
\end{align}
\item With the above specifications the linear map $R:\,^{(4)}\Sigma_t \rightarrow \,^{(4)}\Sigma_t$ is given by
\begin{align}
\label{eq:Rot5DU1}
\left( R(\psi_0)^{\bm}{}_{\bn} \right) = &
\left( \begin{matrix}
1 & 0 & 0 & 0 \\
0 & 1 & 0 & 0 \\
0 & 0 & \cos\psi_0 & -\sin\psi_0 \\
0 & 0 & \sin\psi_0 &  \cos\psi_0 
\end{matrix} \right)
= \left( \begin{matrix}
1 & 0 & 0 & 0 \\
0 & 1 & 0 & 0 \\
0 & 0 & \frac{z}{\rho} & -\frac{w}{\rho} \\
0 & 0 & \frac{w}{\rho} &  \frac{z}{\rho}
\end{matrix} \right)
\,,
\end{align}
where $\psi_0$ is the rotation angle;
\item Now, we compute the numerical data at the grid points $(x,y,z,w=0)$;
\item Afterwards, we have to interpolate the grid functions to points\\
      $p=(x,y,\rho,\psi=0)$;
\item Finally, we generate the function values at grid points $p'=(x,y,z,w\neq0)$
by rotating the data from $p=(x,y,\rho,0)$ by an angle $\psi_0$.
Spatial tensors $T$ 
transform from $p=(x,y,\rho,\psi=0)$ to a grid point $p'=R(p)=(x,y,z,w\neq0)$
according to Eq.~\eqref{eq:TensorTrafo4D} with the rotation matrix now given by Eq.~\eqref{eq:Rot5DU1}.
In particular, scalar, vector and 2-tensor type quantities transform as
\begin{subequations}
\label{eq:Trafo5DU1}
\begin{align}
\Psi(x,y,z,w)   = & \Psi(x,y,\rho,0) \,,\\
V^{\bm}(x,y,z,w)  = & R^{\bm}{}_{\bn} V^{\bn}(x,y,\rho,0) \,,\\
S_{\bm\bn}(x,y,z,w) = & (R^{-1})^{\bk}{}_{\bm} (R^{-1})^{\bl}{}_{\bn} S_{\bk\bl}(x,y,\rho,0) 
\,.
\end{align}
\end{subequations}
\end{enumerate}
Once we have prepared the scheme explicitly, we can implement the Cartoon method 
for the BSSN formulation~\eqref{eq:BSSN}. 

\subsection{Modified Cartoon method in $D>5$}
\label{ssec:ModCartoon}
While the extension of the Cartoon method to $D=5$ spacetime dimensions has been 
straightforward, the generalization to $D\geq6$ dimensions requires
a bit more brain-work.
In particular, we have to constrain ourselves to configurations which exhibit a
$SO(D-3)$ isometry, such that we can reduce them to effectively $3+1$ formulations.
The spatial slice can be represented in
Cartesian coordinates
$(x,y,z,w^{1},\ldots,w^{D-4})$
or curvi-linear coordinates 
$(x,y,\rho,\varphi_{1},\ldots, \varphi_{D-4})$,
where $\rho = \sqrt{z^2 + \sum_{\bmu=1}^{D-4} (w^{\bmu})^{2} }$.
The transformation between the two coordinate systems is given by
\begin{subequations}
\label{eq:CoordTrafoD}
\begin{align}
x = & x\,,\quad
y =   y\,,\\
z = & \rho \cos\varphi_1
\,,\\
w^1 = & \rho \sin\varphi_1 \cos\varphi_2 
\,\,\,\qquad\qquad\qquad\qquad(D\geq6) \,,\\
\vdots \nonumber\\
w^{D-5} = & \rho\sin\varphi_1 \ldots \sin\varphi_{D-5} \cos\varphi_{D-4}
\qquad(D\geq6) \,,\\
w^{D-4} = & \rho \sin\varphi_1 \ldots \sin\varphi_{D-5} \sin\varphi_{D-4}
\,\qquad(D\geq5) 
\,.
\end{align}
\end{subequations}
We choose our symmetries such that the dynamics are confined to the $(x^i)=(x,y,z)$ 
hyperplane and the $SO(D-3)$ symmetry is imposed onto the remaining
$(D-4)$ coordinates $w^{\bmu}$. 
Then, the line element writes
\begin{align}
\label{eq:CartoonDle0}
ds^2 = & - \left( \alpha - \beta^{\bm}\beta_{\bm} \right) dt^2 + 2\beta_{\bm} dt dx^{\bm}
         + \gamma_{ij}(x^{\mu}) dx^{i} dx^{j} + \kappa(x^{\mu}) d\Omega^{2}_{D-4}
\,\nonumber\\ & 
         - \left( \alpha - \beta^{k}\beta_{k} \right) dt^2 + 2\beta_{k} dt dx^{k}
         + \gamma_{ij}(x^{\mu}) dx^{i} dx^{j} + \kappa(x^{\mu}) d\Omega^{2}_{D-4}
\,,
\end{align}
where $d\Omega^{2}_{D-4}$ is the line element of the unit-$(D-4)$ sphere,
the extra-dimensional components $\beta^{\bmu}=0$, 
$\gamma_{ij}$ is the $3$-dimensional spatial metric
and $\kappa=\gamma_{\bmu\bmu}$ can be viewed as a conformal factor for the extra-dimensional
metric components.
Due to the isometry, all geometric quantities are independent of the extra-dimensional coordinates
$w^{\bmu}$ and only
depend on the $4$-dimensional coordinates $x^{\mu}=(t,x^{i})$.
After identifying the necessary symmetries and setting up our coordinates, 
we compute the EoMs~\eqref{eq:BSSN}
in the $(x,y,z,w^{1}=0,\ldots,w^{D-4}=0)$ hyperplane.
In order to evaluate the spatial derivates with respect to 
the extra dimensions
we could \textit{in principle} proceed 
by successively applying the Cartoon method as described in Sec.~\ref{ssec:Cartoon5D}.
However, even this method could become less feasible for a large number of extra dimensions.
Recall that we have to set up additional $(2g+1)$ grid points
for each extra spatial dimension.
Then, the number of grid points would be $N^3 (2g+1)^{(D-4)}$
and therefore the method, too, becomes numerically more expensive with increasing spacetime dimension $D$.

Instead, Shibata \& Yoshino~\cite{Shibata:2010wz}
in their original publication on the \textit{modified Cartoon method}
suggest to directly employ symmetry relations for all extra-dimensional tensor components.
This allows us to re-express all components and their derivatives in the 
$(w^{1},\ldots,w^{D-4})$ directions
by expressions in the $(x,y,z)$ hyperplane.  
Thus, we will again be left with a numerical grid of size $N^3 (2g+1)$
and we can employ the Cartoon method as discussed in Sec.~\ref{ssec:Cartoon5D}.
The beauty of this approach is that the code developed for $D=5$ spacetime dimensions 
can now straightforwardly be generalized to $D\geq6$-dimensional spacetimes
with only little increase in memory requirements
as long as the configuration exhibits a $SO(D-3)$ isometry.

In view of the BSSN equations~\eqref{eq:BSSN} let us focus on the symmetry relations for 
scalars $\Psi$, vectors $V^{\bm}$ and symmetric 2-tensors $S_{\bm\bn}$
in the $(x,y,z)$ hyperplane.
Because of the $SO(D-3)$ isometry we obtain
\begin{subequations}
\label{eq:CartoonDsub1}
\begin{align}
V^{\bmu}  = & 0 
\,,\\
S_{i\bmu} = & 0 \,,\quad
S_{\bmu\bnu} = \delta_{\bmu\bnu} S_{ww}
\,,\\
S_{ww} := & S_{44} = \ldots = S_{D-1 D-1}
\,.
\end{align}
\end{subequations}
The symmetry relations for the various derivatives are given in Eqs.~(29)-(31) 
of Ref.~\cite{Shibata:2010wz}
and I summarize them here for completeness:
\begin{itemize}
\item derivatives of scalars
\begin{align}
\label{eq:CartoonDsubS}
\p_{\bmu} \Psi = & 0
\,,\quad
\p_{\bmu}\p_{\bnu} \Psi = \delta_{\bmu\bnu} \frac{\p_z\Psi}{z}
\,;
\end{align}
\item derivatives of vector components
\begin{subequations}
\label{eq:CartoonDsubV}
\begin{align}
\p_{\bmu} V^{i} = & 0 \,,\quad
\p_{\bmu} V^{\bnu} =   \delta^{\bnu}{}_{\bmu} \frac{V^{z}}{z} 
\,,\quad 
\p_{j}\p_{\bmu} V^{i} = 0 \,,\quad
\p_{\brho}\p_{\bmu} V^{\bnu} = 0
\,,\\
\p_{i}\p_{\bmu} V^{\bnu} = & \delta^{\bnu}{}_{\bmu} \left(\frac{\p_{i}V^z}{z}- \delta_{iz} \frac{V^{z}}{z^2}\right)
\,,\quad
\p_{\bmu}\p_{\bnu} V^{i} = \delta_{\bmu\bnu}\left( \frac{\p_{z}V^{i}}{z} - \delta_{iz} \frac{V^{z}}{z^2}\right)
\,;
\end{align}
\end{subequations}
\item derivatives of tensor components 
\begin{subequations}
\label{eq:CartoonDsubT}
\begin{align}
\p_{\bmu} S_{ij} = & 0 \,,\quad
\p_{\brho} S_{\bmu\bnu} =   0 \,,\quad
\p_{\bmu} S_{i\bnu} = \delta_{\bmu\bnu}\left( \frac{S_{iz}}{z} - \delta_{iz} \frac{S_{ww}}{z} \right)
\,,\\
\label{eq:CartoonDsubT2}
\p_{k}\p_{l} S_{i\bmu}        = & 0\,,\,\,
\p_{k}\p_{l} S_{\bmu\bnu}     =   0\,,\,\,
\p_{k}\p_{\bmu} S_{ij}        =   0\,,\,\,
\p_{k}\p_{\brho} S_{\bmu\bnu} =   0\,,\,\,
\p_{\bnu}\p_{\brho}S_{i\bmu}  =   0
\,,\\
\p_{j}\p_{\bmu} S_{i\bnu} = & \delta_{\bmu\bnu} \left(
        \delta_{jz}\frac{\delta_{iz}S_{ww}-S_{iz} }{z^2} 
        + \frac{\p_{j}S_{iz} -  \delta_{iz}\p_{j}S_{ww} }{z} \right)
\,,\\
\p_{\bmu}\p_{\bnu} S_{ij} = & \delta_{\bmu\bnu}\left( \frac{\p_{z}S_{ij}}{z} - \delta_{iz} \frac{S_{ij}}{z^2}
                        + \delta_{iz}\delta_{jz} \frac{2S_{ww} - S_{zz}}{z^2} \right)
\,,\\
\p_{\brho}\p_{\bsigma}S_{\bmu\bnu} = & \left(\delta_{\bmu\brho}\delta_{\bnu\bsigma} + \delta_{\bmu\bsigma}\delta_{\bnu\brho}\right)
        \frac{S_{zz}-S_{ww}}{z^2}
        + \delta_{\brho\bsigma}\delta_{\bmu\bnu} \frac{\p_{z}S_{ww}}{z}
%
\,;
\end{align}
\end{subequations}
\end{itemize} 
Bear in mind that some of the BSSN variables~\eqref{eq:BSSNvars} are tensor densities,
or derivatives thereof
and it is not immediatly evident that the expressions are valid.
A careful calculation, however, shows that the symmetry relations indeed cary over to the conformal factor, metric
and tracefree part of the extrinsic curvature.
Note, that there are some terms in Eqs.~\eqref{eq:CartoonDsubS} -~\eqref{eq:CartoonDsubT}
which behave as $\tfrac{1}{z}$ or $\tfrac{1}{z^2}$.
Although these terms are perfectly regular analytically, a numerical scheme will 
diverge at $z=0$ due to explicit division by $0$.
Therefore, we have to enforce the regularity of these terms by
substituting them with regular expressions in a neighbourhood 
of $z=0$~\footnote{Note, that we will encounter similar challenges for the \textit{dimensional reduction}
approach presented in the following section~\ref{sec:DimRed}}.  
To give you a flavour of the \textit{regularization} procedure 
let us consider a function $f=f(x,y,z)$ which
is linear in $z$ near the axis. Then, we can Taylor expand this function around $z=0$
according to $f = z \p_{z} f|_{z=0} + \O(z^2) = z f_{1} + \O(z^2)$. 
In the limit $z\rightarrow0$ we obtain
\begin{align}
\label{eq:Reg1}
\lim_{z\rightarrow0} \frac{f(x,y,z)}{z} = & f_{1} = \p_z f(x,y,z)|_{z=0}
\,.
\end{align}
In a similar manner we can regularize all the terms in question.
Instead of presenting the relations here I refer to the 
list of substitution rules given in Eq.~(32) of Ref.~\cite{Shibata:2010wz}
or in App.~B of Ref.~\cite{Zilhao:2010sr}.
In practise, we replace terms $\sim\tfrac{1}{z}$ or $\sim\tfrac{1}{z^2}$ in our implementation
with these regularized expression whenever it has to be evaluated at or close to $z=0$.

The modified Cartoon method for higher dimensional spacetimes has proven to be a 
very robust numerical method. 
For example, it has been employed to explore singly spinning Myers-Perry 
BHs in $D=5,\ldots8$~\cite{Shibata:2010wz}.

\newpage
\subsection{Exercises}\label{ssec:ExerciseCartoon}
\subsubsection{Cartoon method in $D=5$ with $U(1)\times U(1)$ symmetry}
\label{sssec:Cartoon2x}
Develop the Cartoon method for the doubly spinning Myers-Perry BH in $D=5$
which allows to reduce the problem to a $2+1$-dimensional setup.
The Myers-Perry BH with $N=\left[\frac{D-1}{2}\right]$ independent rotation planes 
in odd-dimensional spacetimes~\footnote{In even dimensions there is 
an additional term $r^2 d\alpha^2$ such that $\sum_{i=1}^N \mu_i^2 + \alpha^2=1$~\cite{Emparan:2008eg}.}
is given by~\cite{Emparan:2008eg,Myers:1986un}
\begin{align}
\label{eq:MP5D2a}
ds^2 = & -d t^2 + \frac{F P}{P - r^{D-3}_S r^2}dr^2
\nonumber\\ &
        + \sum_{i=1}^{N} \left(r^2 + a_i^2\right) \left( d\mu_i^2 + \mu_i^2 d\varphi^2\right)
                        +\frac{r^{D-3}_S r^2}{F P}\left( dt - a_i \mu_i^2 d\varphi^2\right)^2
\,,
\end{align}
where $a_i$  (with $i=1,\ldots,N$) are the spin parameters,
$r_S$ is the horizon radius, 
$r^{D-3}_S = \frac{16 \pi M}{(D-2)\mathcal{A}_{D-2}}$ denotes the mass parameter,
$\mu_i$ are directional cosines, i.e., $\sum_{i=1}^N\mu_i^2=1$,
and the functions $F$ and $P$ are
\begin{align}
F = & 1-\sum_{i=1}^{N}\frac{a_i^2 \mu_i^2}{r^2+a_i^2}
\,,\qquad
P =  \prod_{i=1}^N \left(r^2 + a_i^2\right) 
\,.
\end{align}
%

\subsubsection{Cartoon method in $D\geq6$} 
\label{sssec:Cartoon6D}
Verify the symmetry relations~\eqref{eq:CartoonDsubS}--~\eqref{eq:CartoonDsubT}
for the derivatives of the BSSN variables~\eqref{eq:BSSNvars}.
Recall, that the BSSN variables $(\chi,\tg_{ij},\tA_{ij})$ are tensor densities, i.e., 
\begin{align}
\tilde{T} = & \gamma^{\frac{\mathcal{W}}{2}} T
\,,
\end{align}
where $\mathcal{W}$ is the tensor weight. Instead $\tG^{i}$ is the derivative of the 
tensor density $\tg^{ij}$.


\newpage
\section{Dimensional reduction and effective $3+1$ formulation}\label{sec:DimRed}
In this section I present a second approach to simulate higher dimensional BH spacetimes
based on the \textit{dimensional reduction by isometry}.
This method has been applied successfully to numerically evolve head-on collisions 
of BHs in $D=5$ dimensions~\cite{Zilhao:2010sr,Witek:2010xi,Witek:2010az}
and $D=6$ dimensions~\cite{Witek:2013koa,HigherDWiP}
The dimensional reduction is a well developed concept in theoretical physics.
For example, it has been proposed to unify the Einstein-Maxwell theory in $D=4$
as a $5$-dimensional (pure) gravity theory as first introduced by
Kaluza~\cite{Kaluza:1921tu} and Klein~\cite{Klein:1926tv}.
Later, the Kaluza-Klein (KK) reduction has been used to unify gravity with more general
gauge theories and KK-BHs have attracted a lot of 
attention~\cite{Harmark:2005pp,Myers:1986rx,Gibbons:1985ac}.
Recently, the KK compactification has been used to develop a map between AdS and Ricci-flat spacetimes 
which, in turn, has been applied to investigate, e.g., the Gregory-Laflamme instability~\cite{Caldarelli:2012hy}.
Conversely, a higher dimensional gravity theory can be formulated as a lower dimensional one
but coupled to gauge and scalar fields.
The original KK reduction is, in fact, a compactification over a compact manifold
and the reduced theory is regarded as a low-energy approximation obtained by 
keeping only the zero KK-modes.

Here, instead, I present a Killing reduction formalism,
where the reduction is \textit{not} an approximation but follows directly from the isometry.
This idea dates back to Geroch's work in $D=4$~\cite{Geroch:1970nt}
which has been applied to numerical evolutions, e.g., in Ref.~\cite{Rinne:2005sk}
Geroch's formalism has later been extended to higher dimensions~\cite{Chiang:1985rk, Cho:1986wk,Cho:1987jf}.
A further, very pedagogical discussion of the subject can be found in Zilh\~{a}o~\cite{Zilhao:2013gu}.

To grasp the basic concept let us consider a $D$-dimensional 
manifold $(\M,g_{MN})$ which exhibits KV fields $\xi^a$
that are everywhere either timelike or spacelike.
The collection $S$ of all integral curves of $\xi^a$ forms a lower dimensional 
quotient space of $\M$.
If tensor fields $T^{M_1\ldots M_p}{}_{N_1\ldots N_q} \in \M$ satisfy
\begin{subequations}
\label{eq:DimRedSym}
\begin{align}
\xi^{A} T^{M_1\ldots M_p}{}_{A N_2\ldots N_q} = & 0 \,,\ldots\,,
\xi_{A} T^{M_1\ldots M_{p-1}A}_{N_1\ldots N_q} = 0 
\,,\quad\text{and}\\
\Lie_{\xi} T^{M_1\ldots M_p}{}_{N_1\ldots N_q} = & 0
\,,
\end{align}
\end{subequations}
then one can show~\cite{Geroch:1970nt} 
that there is a one-to-one mapping between tensor fields $T^{M_1\ldots M_p}{}_{N_1\ldots N_q} \in \M$
and $\tilde{T}^{M_1\ldots M_p}{}_{N_1\ldots N_q} \in S$.
In other words, the entire tensor field algebra in $S$ is uniquely and completely determined  
by tensors $T^{M_1\ldots M_p}{}_{N_1\ldots N_q} \in \M$ that satisfy Eqs.~\eqref{eq:DimRedSym}.
We denote the metric on $S$ as $h_{MN}$ and it is related to the $D$-dimensional one via
\begin{align}
h_{MN} = & g_{MN} - \frac{1}{\lambda} \xi_{M} \xi_{N}
\,,\quad
h^{MN} =   g^{MN} - \frac{1}{\lambda} \xi^{M} \xi^{N}
\,,
\end{align}
where $\lambda = \xi^{M} \xi_{M}$ is the norm of the KV fields.
This relation immediately provides a projection operator onto S
\begin{align}
h^{M}{}_{N} = & \delta^{M}{}_{N} - \frac{1}{\lambda} \xi^{M} \xi_{N}
\,.
\end{align}
Based on these relations and the above symmetries one can show~\cite{Geroch:1970nt,Chiang:1985rk}
that pure (vacuum) gravity in $D$ dimensions is indeed equivalent to 
a lower dimensional gravity theory coupled to gauge and scalar fields.

\subsection{Dimensional reduction by isometry and $(D-4)+4$ split}
\begin{figure}[htpb!]
\begin{center}
\includegraphics[width=0.7\textwidth]{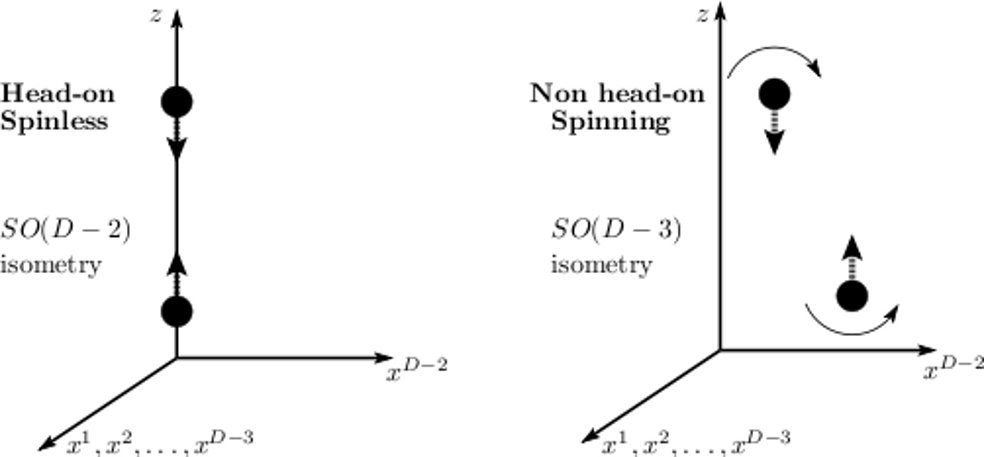}
\end{center}
\vspace*{8pt}
\caption{\label{fig:DimRedSym}
Illustration of classes of models accessible by the dimensional reduction on a $(D-4)$ sphere.
Left: models in $D\geq5$ with an $SO(D-2)$ isometry, e.g., head-on collisions of non-rotating BHs;
Right: models in $D\geq6$ with an $SO(D-3)$ isometry, e.g., collisions of BHs with an impact parameter
or spinning BHs as long as the dynamics are confined to one plane.
Taken from~\cite{Zilhao:2010sr}.
}
\end{figure}

Here, I will derive the formalism based on the dimensional reduction 
focusing on the isometries that are relevant for numerical simulations.
As I have discussed before, we desire to reduce our problems to an effectively
$3+1$-dimensional one, 
which allows us to generalize existing NR codes in a straightforward manner.
In order to end up with a $4$-dimensional base space, we have to perform
the dimensional reduction on a $(D-4)$-sphere, 
which implies a $SO(D-3)\subset SO(D-2)$ isometry group.
Due to this isometry, there are $N=\tfrac{(D-4)(D-3)}{2}$ independent KV
fields $\xi^a$ (with $a=1,\ldots,N$) which satisfy the Lie algebra
\begin{align}
\label{eq:DimRedLieAl}
\left[\xi_a,\xi_b \right] = & \epsilon_{ab}{}^{c}\xi_{c}
\,,
\end{align}
and $\epsilon_{ab}{}^{c}$ are the structure constants of the $SO(D-3)$.
Then, as illustrated in Fig.~\ref{fig:DimRedSym},
the possible classes of models that we will be able to explore include:
\begin{itemize}
\item in $D\geq5$: configurations that exhibit a $SO(D-2)$ symmetry, i.e.,
                   axissymmetric setups, such as head-on collisions of non-spinning BHs;
\item in $D\geq6$: configurations that exhibit a $SO(D-3)$ symmetry, such as
  \begin{itemize}
  \item BH collisions with an impact parameter,
  \item rotating BHs with the spin being orthogonal to one plane.
  \end{itemize}
\end{itemize}

The most general ansatz for the $D$-dimensional metric can be written as
\begin{align}
\label{eq:DimRedDmetric}
ds^2 = &  g_{MN} dx^{M} dx^{N}
\nonumber \\ = &
          g_{\mu\nu}(x^M) dx^{\mu} dx^{\nu} 
        + \Omega_{\bmu\bnu}(x^M)\left( dx^{\bmu} - A^{\bmu}{}_{\mu}(x^M) dx^{\mu} \right)
                                \left( dx^{\bnu} - A^{\bnu}{}_{\nu}(x^N) dx^{\nu} \right)
\nonumber \\ = &
          \left( g_{\mu\nu} + \Omega_{\bmu\bnu} A^{\bmu}{}_{\mu} A^{\bnu}{}_{\nu} \right) dx^{\mu} dx^{\nu}
        - 2 \Omega_{\bmu\bnu} A^{\bmu}{}_{\mu} dx^{\mu} dx^{\bnu}
        + \Omega_{\bmu\bnu} dx^{\bmu} dx^{\bnu}
\,,
\end{align}
where $M,N=0,\ldots,D-1$ are the $D$-dimensional spacetime indices,
$\mu,\nu=0,\ldots,3$ refer to the $4$-dimensional base spacetime
and $\bmu,\bnu=4,\ldots,D-1$ denote the extra dimensions.
In other words, we can view the $D$-dimensional spacetime as a fibre bundle with
coordinates $\{x^{\mu}\}$ in the base space and
coordinates $\{x^{\bmu}\}$ along the fibre.

Bearing in mind that the fibre has the minimal dimension necessary
to accommodate the $\tfrac{(D-4)(D-3)}{2}$ independent KV fields $\xi^{a}$
we may assume without loss of generality that the KVs have components
exclusively along the fibre
and normalize them, such that they only depend on the coordinates of the fibre, i.e.,
\begin{align}
\xi_a = & \xi^{\bmu}_{a} \p_{\bmu}
\,,\qquad\text{and}\qquad
\p_{\mu} \xi^{\bmu}_{a} = 0
\,.
\end{align}
Using the fact that $\xi^{a}$ are KVs of the spacetime, i.e., that they satisfy
\begin{align}
\Lie_{\xi} g_{MN} = & 0
\,,
\end{align}
we can derive properties of the metric functions 
appearing in Eq.~\eqref{eq:DimRedDmetric}.
In particular we obtain
\begin{enumerate}
\item for $M=\bmu,N=\bnu$: 
\begin{align}
g_{MN} = & \Omega_{\bmu\bnu} 
\qquad\Rightarrow\qquad
\Lie_{\xi}\Omega_{\bmu\bnu} = 0
\,.
\end{align}
This relation implies that $\Omega_{\bmu\bnu}$ admits the maximal number of KVs and, therefore, is the metric on the maximally
symmetric space at each $x^{\mu}$.
From the commutation relation~\eqref{eq:DimRedLieAl} it follows that this maximally symmetric space is the 
$S^{D-4}$-sphere. Therefore, we obtain
\begin{align}
\label{eq:DimRedRel1}
\Omega_{\bmu\bnu} = & \lambda(x^{\mu}) h^{S^{D-4}}_{\bmu\bnu}
\,,
\end{align}
where $h^{S^{D-4}}_{\bmu\bnu}$ denotes the metric on the unit $S^{D-4}$-sphere.
Remember, that the scalar function $\lambda(x^{\mu})$ is the norm of the KV fields.
\item for $M=\mu,N=\bnu$: 
\begin{align}
g_{MN} = &\Omega_{\bmu\bnu}A^{\bmu}{}_{\mu}
\qquad\Rightarrow\qquad
\Lie_{\xi} A^{\bmu}{}_{\mu} = 0
\,.
\end{align}
This relation is equivalent to $\left[\xi_a,A_{\mu}\right]=0$  and implies
\begin{align}
\label{eq:DimRedRel2}
A^{\bmu}{}_{\mu} = & 0
\,,
\end{align}
which means that for $D\geq5$ there are no non-vanishing vector fields
on the $S^{D-4}$-sphere which commute with all KVs on the sphere.
Another way of interpreting Eq.~\eqref{eq:DimRedRel2} is that
all KVs $\xi^a$ must be hypersurface orthogonal.
In group theory language this relation corresponds to the fact that the gauge group for a theory
reduced on a coset space $G/H$ is the normalizer of $H$ in $G$~\cite{Cho:1986wk,Cho:1987jf}.
In the present case of a sphere, this normalizer (or gauge group) vanishes, i.e., $g_{\mu\bmu}=0$ and there
are no gauge vector fields. 
\item for $M=\mu,N=\nu$: 
\begin{align}
g_{MN} = & g_{\mu\nu} + \Omega_{\bmu\bnu}A^{\bmu}{}_{\mu} A^{\bnu}{}_{\nu}
\qquad\Rightarrow\qquad
\Lie_{\xi} g_{\mu\nu} = 0
\,.
\end{align}
This relation tells us that the KVs act transitively on the fibre implying that 
the base space metric is independent of the fibre coordinates, i.e., 
\begin{align}
\label{eq:DimRedRel3}
g_{\mu\nu}(x^M) = & g_{\mu\nu} (x^{\mu})
\,.
\end{align}
\end{enumerate}
The properties~\eqref{eq:DimRedRel1},~\eqref{eq:DimRedRel2} and~\eqref{eq:DimRedRel3}
imply that the $D$-dimensional metric $g_{MN}$ 
generally given by Eq.~\eqref{eq:DimRedDmetric} 
now has a block diagonal form 
\begin{align}
\label{eq:DimRedMetric}
d s^2 = g_{MN} dx^M dx^N 
      = g_{\mu\nu}(x^{\mu}) dx^{\mu} dx^{\nu} & + \lambda(x^{\mu}) h^{S^{D-4}}_{\bmu\bnu} dx^{\bmu} dx^{\bnu}
\nonumber \\     
      = g_{\mu\nu}(x^{\mu}) dx^{\mu} dx^{\nu} & + \lambda(x^{\mu}) d\Omega^2_{D-4}
\,, \\ 
 D=4 \quad & \qquad\quad D-4
\nonumber
\end{align}
where the components of the base- and fibre-space, respectively, are clearly separated.
In other words, after performing the dimensional reduction on a $S^{D-4}$ sphere
our $D$-dimensional vacuum spacetime is uniquely described by the $4$-dimensional metric $g_{\mu\nu}(x^{\mu})$
and the scalar field $\lambda(x^{\mu})$, depending only on coordinates of the
$4$-dimensional base space.
Note, however, that $\lambda(x^{\mu})$ behaves as a radial-like function 
and, in particular, the authors of Ref.~\cite{Zilhao:2010sr} have chosen 
\begin{align}
\label{eq:LamY2}
\lambda(x^{\mu})\sim g_{\theta\theta} \sin^2\theta \sim y^2
\,.
\end{align}
These relations become equalities for axissymmetric configurations and in coordinates adapted to this axial symmetry.
In order to derive the EoMs, we perform the $(D-4)+4$ split of 
the Ricci tensor $\,^{(D)}R_{MN}$. Using Eq.~\eqref{eq:DimRedMetric}, the various components become
\begin{subequations}
\label{eq:DimRedDRicci}
\begin{align}
M=\bmu\ N=\bnu: & \,^{(D)}R_{\bmu\bnu} = \,^{(D-4)}R_{\bmu\bnu} 
                - \frac{1}{2} h^{S^{D-4}}_{\bmu\bnu} \left( \nabla^{\mu}\nabla_{\mu}\lambda + \frac{D-6}{2\lambda} \nabla^{\mu}\lambda \nabla_{\mu}\lambda \right)
\,,\\
M=\mu, N=\bmu: & \,^{(D)}R_{\mu\bmu} = 0
\,,\\
M=\mu, N=\nu: & \,^{(D)}R_{\mu\nu} = \,^{(4)}R_{\mu\nu} 
                - \frac{D-4}{2\lambda}\left( \nabla_{\mu}\nabla_{\nu}\lambda -\frac{1}{2\lambda} \nabla_{\mu}\lambda\nabla_{\nu}\lambda \right)
\,,
\end{align}
\end{subequations}
where $\nabla_{\mu}$ is the covariant derivative with respect to the $4$-metric and the 
extra dimensional Ricci tensor is
\begin{align}
\,^{(D-4)}R_{\bmu\bnu} = & (D-5) h^{S^{D-4}}_{\bmu\bnu}
\,.
\end{align}
We are interested in $D$-dimensional vacuum spacetimes and therefore $\,^{(D)}R_{MN}=0$.
Then, using Eqs.~\eqref{eq:DimRedDRicci}, we obtain the EoM for the scalar field
as well as Einstein's equations non-minimally coupled to the scalar field
\begin{subequations}
\label{eq:DimRedEoMs}
\begin{align}
\label{eq:DimRedEoMSc}
\S = & \nabla^{\mu}\nabla_{\mu} \lambda + \frac{D-6}{2\lambda}\nabla^{\mu}\lambda\nabla_{\mu}\lambda - 2(D-5) = 0
\,,\\
\label{eq:DimRedEoMGR}
E_{1,\mu\nu} = & \,^{(4)}R_{\mu\nu} -\frac{1}{2}g_{\mu\nu} \,^{(4)}R - 8 \pi \,^{(4)}T_{\mu\nu} = 0
\,,\quad\text{with}\\
\label{eq:DimRedEoMTmn}
\,^{(4)}T_{\mu\nu} = & \frac{D-4}{16\pi\lambda} 
                       \left(  \nabla_{\mu}\nabla_{\nu}\lambda 
                             - \frac{1}{2\lambda}\nabla_{\mu}\lambda\nabla_{\nu}\lambda
                             - (D-5) g_{\mu\nu}
                             + \frac{D-5}{4\lambda} g_{\mu\nu} \nabla^{\kappa}\lambda \nabla_{\kappa}\lambda \right)
\,.
\end{align}
\end{subequations}
We observe, that our dynamics now really only depend on $4$-dimensional gravity, coupled to a scalar field $\lambda$ which 
encodes all the information about the extra dimensions~\footnote{Eqs.~\eqref{eq:DimRedEoMs} only 
depend on $4$-dimensional quantities and we will therefore drop all dimensional superscripts
for the reminder of this section.}.

\subsection{$3+1$ decomposition and BSSN formulation}
Let me repeat my statement at the end of last section:
By means of the dimensional reduction on a $S^{D-4}$-sphere we
have formulated the $D$-dimensional vacuum Einstein's equations as $4$-dimensional Einstein's equations
coupled (non-minimally) to a scalar field. 
This observation is quite significant and allows us to numerically investigate higher dimensional gravity
by evolving the modified equations on a $3+1$-dimensional domain.
For this purpose we need to re-write the dimensionally reduced EoMs~\eqref{eq:DimRedEoMs} 
as time evolution problem as outlined in Sec.~\ref{sec:Decomp}.
Because our base space is $4$-dimensional we will replace $D\rightarrow\tilde{D}=4$.

The dynamical variables are the spatial metric $\gamma_{ij}$ on the $3$-dimensional spatial slice,
the corresponding extrinsic curvature $K_{ij}$, the scalar field $\lambda$ and
its conjugated momentum $K_{\lambda}$ 
\begin{align}
\label{eq:DimRedKl}
K_{\lambda} = & -\frac{1}{2\alpha} \left(\p_t-\Lie_{\beta}\right) \lambda
\,,
\end{align}
which we introduce to close the system.
The scalar sector of the EoMs is derived from Eqs.~\eqref{eq:DimRedEoMSc} and~\eqref{eq:DimRedKl}
\begin{subequations}
\label{eq:DimRedADMs}
\begin{align}
\label{eq:DimReddtLam}
\p_{t}\lambda     = & -2\alpha K_{\lambda} + \Lie_{\beta}\lambda
\,,\\
\p_{t}K_{\lambda} = &  \alpha \left(  (D - 5) - \frac{1}{2} D^{i}D_{i}\lambda 
               - \frac{D - 6}{4\lambda} D^{i}\lambda D_{i}\lambda
               + K K_{\lambda} 
               + \frac{D-6}{\lambda} K_{\lambda}^2 \right)
\nonumber \\ &
        - \frac{1}{2} D^{i}\alpha D_{i}\lambda  
        + \mathcal{L}_{\beta} K_{\lambda}
\label{eq:DimReddtKLam}
\end{align}
\end{subequations}
The tensor sector of EoMs is described by the ADM-like equations~\eqref{eq:ADMeqs},
where the energy density, flux and spatial part are determined by the various 
projections~\eqref{eq:ProjTmn} of the energy-momentum tensor given in Eq.~\eqref{eq:DimRedEoMTmn}
with respect to the now $3$-dimensional spatial slice.
The evolution equations for the $3$-metric and extrinsic curvature
are then modified to
\begin{subequations}
\label{eq:DimRedADMg}
\begin{align}
\label{eq:DimReddtMetric}
\p_t \gamma_{ij} = & -2\alpha K_{ij} + \Lie_{\beta} \gamma_{ij}
\,,\\
\label{eq:DimReddtExCurv}
\p_t K_{ij} = & -D_{i}D_{j}\alpha + \alpha \left( R_{ij} + K K_{ij} - 2 K^{k}{}_{i} K_{kj} \right)
                +\Lie_{\beta} K_{ij}
        \nonumber\\ &
                - \alpha\frac{D-4}{2\lambda} \left( D_i D_j\lambda - 2 K_{\lambda}K_{ij} 
                        - \frac{1}{2\lambda}D_i\lambda D_j\lambda \right)
\,,
\end{align}
\end{subequations}
where the scalar field enters with second derivatives, thus changing the principal symbol,
i.e., the character of the PDE system.
The physical constraints become
\begin{subequations}
\label{eq:DimRedADMc}
\begin{align}
\H = & R - K_{ij} K^{ij} + K^2 \nonumber \\ &
       - \frac{D-4}{\lambda}\left(  D^{i}D_{i}\lambda +\frac{D-7}{4\lambda} D^{i}\lambda D_{i}\lambda
                                  - (D-5)\frac{K_{\lambda}^2}{\lambda} - 2 K K_{\lambda} - (D-5) \right)
\,,\\
\M_i = & D_{j}K^{j}{}_{i} - D_{i} K 
        -\frac{D-4}{2\lambda} \left( 2 D_i K_{\lambda} -\frac{K_{\lambda}}{\lambda} D_i\lambda - K^{j}{}_{i}D_{j}\lambda \right)
\,,
\end{align}
\end{subequations}
which need to be solved for the initial data. For a discussion of the initial data construction
in general I refer to Refs.~\cite{Cook:2000vr, Okawa:2013afa}
and to Refs.~\cite{Zilhao:2010sr,Zilhao:2011yc} 
for the present (dimensionally reduced) system.

The evolution equations~\eqref{eq:DimRedADMs} and~\eqref{eq:DimRedADMg} are still in ADM-like form
and therefore only weakly hyperbolic. 
As we have noted before and is discussed in detail in Hilditch's contibution to the lecture notes~\cite{Hilditch:2013sba}
this formulation is prone to numerical instabilities due to its PDE structure.
To ``cure'' this instability we need to re-write Eqs.~\eqref{eq:DimRedADMs} and~\eqref{eq:DimRedADMg}
in a strongly hyperbolic formulation.
Therefore, we cast them in the well-established BSSN scheme 
discussed in Sec.~\ref{ssec:DecompEE}.
Again, we change the dynamical variables by introducing the trace and trace-free part of 
the extrinsic curvature, the conformal connection function and conformally decomposing the metric.
In addition, we have to rescale the additional scalar field $\lambda\sim y^2$ 
(see Eq.~\eqref{eq:LamY2})
to make its coordinate dependence explicit and, thus, allow for a straight-forward regularization of the
variable.
Bearing in mind that our computational domain is $\tilde{D}=3+1$ dimensional and substituting
$D\rightarrow \tilde{D}=4$ in Eqs.~\eqref{eq:BSSNvars}
the new, independent variables are~\footnote{Note, that I define 
the re-scaled variable $K_{\zeta}\sim K_{\lambda}$
differently from the original paper~\cite{Zilhao:2010sr}.
My choice has proven to yield numerically stable evolutions in $D=6$
spacetime dimensions (see Ref.~\cite{Witek:2013koa}).
}
\begin{subequations}
\label{eq:DimRedBSSNvars}
\begin{align}
\label{eq:DimRedBSSNmetric}
\chi = & \gamma^{-\frac{1}{3}}
\,,\qquad 
\tg_{ij} = \gamma^{-\frac{1}{3}} \gamma_{ij} = \chi \gamma_{ij}
\,,\\
\label{eq:DimRedBSSNKA}
K = & \gamma^{ij} K_{ij}
\,,\quad
\tA_{ij} = \chi A_{ij} = \chi\left( K_{ij} - \frac{1}{3} \gamma_{ij}K \right)
\,,\\
\label{eq:DimRedBSSNGam}
\tG^{i} = & \tg^{kl} \tG^{i}{}_{kl} = - \p_{j} \tg^{ij}
\,,\\
\label{eq:DimRedBSSNSca}
\zeta     = & \frac{\chi}{y^2}\lambda
\,,\qquad
K_{\zeta} =   \frac{\chi}{y^2}K_{\lambda}
\end{align}
\end{subequations}
with $\gamma=\det\gamma_{ij}$ and $\tg = \det\tg_{ij} =1$. 
As before, these definitions give rise to additional algebraic and differential constraints
\begin{align}
\D = & \ln(\tg) = 0
\,,\qquad
\T =   \tg^{ij}\tA_{ij} = 0
\,,\\
G_{i} = & \tg_{ij} \tG^{j} - \tG^{\rm{[DEF]}}_{i} = \tg_{ij} \tG^{j} - \tg^{kl}\p_{k}\tg_{il} = 0
\,.
\end{align}
In order to obtain the strongly hyperbolic BSSN form of the time evolution equations 
we have to modify the PDE structure by adding the constraints
according to Eqs.~\eqref{eq:BSSNevol0}.
However, now the ``[ADM]'' and the constraints in Eqs.~\eqref{eq:BSSNevol0} refer to the 
dimensionally reduced version, Eqs.~\eqref{eq:DimRedADMg} and~\eqref{eq:DimRedADMc}.
Performing both the constraint addition and change of variables yields the BSSN equations~\eqref{eq:BSSN}
with the energy momentum tensor~\eqref{eq:DimRedEoMTmn}
enlargened by the additional evolution equations for $\zeta$ and $K_{\zeta}$.
For sake of completeness, let me write down the time evolution equations in all their beauty
\begin{subequations}
\label{eq:DimRedBSSNev}
\begin{align}
\p_t \chi       = & [\textrm{BSSN}]
\,,\\
\p_t \tg_{ij}   = & [\textrm{BSSN}]
\,,\\
\p_t K          = & [\textrm{BSSN}] + \alpha (D-4) S^{K} 
\,,\\
\p_t \tA_{ij}   = & [\textrm{BSSN}] + \alpha (D-4) \chi S^{\tA}_{ij}
\,,\\
\p_t \tG^i      = & [\textrm{BSSN}] + \alpha (D-4) \tg^{ij} S^{\tG}_j
\,,\\
\p_t \zeta      = &  -2\alpha \left( K_{\zeta} - \zeta \frac{K}{3} \right)
                     + 2\zeta\frac{\beta^{y}}{y} + \beta^{i}\p_i\zeta - \frac{2}{3}\zeta\p_i\beta^i
\,,\\
\p_t K_{\zeta}  = & \beta^{i}\p_i K_{\zeta} - \frac{2}{3} K_{\zeta} \p_i\beta^i
                + \frac{\alpha}{2} \left( \zeta \tD^i \tD_i\chi - \chi \tD^i \tD_i\zeta \right)
                + \frac{1}{2} \tD^i\alpha (\zeta \tD_i\chi - \chi \tD_i\zeta )
\nonumber\\ &
                -\frac{\alpha}{4}\left(  (D-6) \frac{\chi}{\zeta} \tD^{i}\zeta \tD_{i}\zeta
                                       - (2D-7) \tD^i\chi \tD_i\zeta
                                       + (D-1) \frac{\zeta}{\chi}\tD^i\chi \tD_i\chi \right)
\nonumber\\ &
                +\alpha K_{\zeta}\left( \frac{5K}{3} + (D-6) \frac{K_{\zeta}}{\zeta} \right)
                - (D-5)\alpha\chi\frac{\zeta \tg^{yy} - 1}{y^2}
                + 2 K_{\zeta} \frac{\beta^y}{y} 
                + \alpha\chi\zeta \frac{\tG^y}{y}
\nonumber\\ &
                - \chi \zeta \frac{\p^y\alpha}{y}
                - (D-4)\alpha\chi \frac{\p^y\zeta}{y} 
                + \frac{(2D-7)\alpha\zeta}{2} \frac{\p^y\chi}{y} 
\,,
\end{align}
\end{subequations}
where ``[BSSN]'' denotes the BSSN Eqs.~\eqref{eq:BSSN} with $D\rightarrow\tD=4$.
The coupling terms $S^{K}$, $S^{\tA}_{ij}$ and $S^{\tG}_j$ are given by
%
\begin{subequations}
\label{eq:HDFgBSSNcoupling}
\begin{align}
S^{K} = &  (D-5) \frac{\chi}{\zeta} \frac{\zeta\tg^{yy} - 1}{y^2} 
                 - (D-5) \frac{K_{\zeta}^2}{\zeta^2} 
                 - K \frac{K_{\zeta}}{\zeta}
                 + \frac{1}{2}\left( \frac{\chi}{\zeta} \tD^i\tD_i\zeta - \tD^i \tD_i\chi \right)
\nonumber\\ &
                 + \frac{D-1}{4\chi} \tD^i\chi\tD_i\chi 
                 + \frac{(D-6)\chi}{4\zeta^2} \tD^i\zeta \tD_i\zeta
                 - \frac{2D-7}{4\zeta} \tD^i\chi \tD_i\zeta
\nonumber\\ &
                 - \frac{2D-7}{2} \frac{\p^y\chi}{y}
                 + \frac{(D-4)\chi}{\zeta} \frac{\p^y\zeta}{y}
                 - \chi \frac{\tG^y}{y} 
\,,\\
S^{\tA}_{ij} = &  \tA_{ij} \frac{K_{\zeta}}{\chi\zeta}
                + \frac{1}{2\chi} [\tD_i\tD_j\chi]^{\rm{tf}}
                - \frac{1}{2\zeta} [\tD_i\tD_j\zeta]^{\rm{tf}}
                - \frac{1}{4\chi^2} [\p_i\chi\p_j\chi]^{\rm{tf}}
                + \frac{1}{4\zeta^2} [\p_i\zeta \p_j\zeta]^{\rm{tf}}
\nonumber\\ &
                + \frac{1}{2\zeta y} \big( 2\zeta\tG^y_{ij} 
                        - \delta^y_i\p_j\zeta - \delta^y_j\p_i\zeta \big)
                + \frac{\tg_{ij}}{3\zeta} \frac{\p^y\zeta}{y}
                - \frac{\tg_{ij}}{3} \frac{\tG^y}{y}
\,,\\
S^{\tG}_i = & - \frac{2}{\zeta} \p_i K_{\zeta}
                 + \left( \frac{K_{\zeta}}{\zeta} - \frac{K}{3} \right) \frac{\p_i\chi}{\chi}
                 + \left( \frac{K_{\zeta}}{\zeta} + \frac{K}{3} \right) \frac{\p_i\zeta}{\zeta}
\nonumber\\ &
                 + \tA^k_i \left( \frac{1}{\zeta} \p_k \zeta - \frac{1}{\chi}\p_k\chi \right)   
                 + \frac{2}{y}\left( \tA^y_i 
                        + \delta^y_i \left( \frac{K}{3} -\frac{K_{\zeta}}{\zeta}\right) \right)
\,.
\end{align}
\end{subequations}
Notice, that we encounter terms $\sim\tfrac{1}{y}$ or $\sim\tfrac{1}{y^2}$ which might cause numerical divergences
at and close to $y=0$. Therefore, we have to regularize them in a similar manner as discussed in
Sec.~\ref{ssec:ModCartoon} substituting $z\rightarrow y$ in Eq.~\eqref{eq:Reg1}.

We close the system by choosing appropriate gauge conditions for the lapse function $\alpha$ and shift
vector $\beta^{i}$. Specifically, we use a modified version of the puncture gauge~\eqref{eq:MovingPuncture}
in which we account for the contribution by the scalar field $\zeta$ and its momentum $K_{\zeta}$.
In terms of the BSSN variables the modified ``1+log''-slicing and $\Gamma$-driver shift condition write
\begin{subequations}
\label{eq:DimRedGauge}
\begin{align}
\label{eq:DimRed1log}
\p_t \alpha = & \beta^{i}\p_{i}\alpha - 2\alpha\left( K + (D-4)\mu_{\zeta}\frac{K_{\zeta}}{\zeta}\right)
\,,\\
\label{eq:DimRedGD}
\p_t \beta^{i} = &  \beta^{k}\p_{k}\beta^{i} - \eta_{\beta} \beta^{i} + \eta_{\Gamma}\tG^{i} 
                    + \eta_{\zeta} \frac{D-4}{2} \frac{\tg^{ij}\p_{j} \zeta}{\zeta}
\,,
\end{align}
\end{subequations}
where, typically, $\mu_{\zeta}=1$ has proven to be a reasonable choice yielding long-term stable numerical evolution.
The picture is different when it comes to the parameters in the shift condition and
the particular choice of the coefficients $(\eta_{\beta},\eta_{\Gamma},\eta_{\zeta})$ appears to depend on
the setup and dimensionality in a non-trivial way.

The \textit{dimensional reduction by isometry} method has been employed successfully 
to explore head-on collisions in higher dimensions and to compute the first 
gravitational wave signals emitted during the merger and plunge
in $D=5$~\cite{Witek:2010xi,Witek:2010az} and in $D=6$~\cite{Witek:2013koa,HigherDWiP}.

\newpage
\subsection{Exercises}
\subsubsection{ADM form of the modified Einstein's equations}
\label{ssec:DimRedADM}
Derive the ADM form~\eqref{eq:DimRedADMs},~\eqref{eq:DimRedADMg} and~\eqref{eq:DimRedADMc}
of the modified Einstein's equations~\eqref{eq:DimRedEoMs}
obtained from the dimensional reduction by isometry.
For this purpose you can modify the \textsc{mathematica} notebook ``GR\_Split.nb'' available online~\cite{ConfWeb}.

\subsubsection{Regularization of terms $\sim \tfrac{1}{y}$ and $~\sim \tfrac{1}{y^2}$}
\label{ssec:Regularization}
The BSSN evolution Eqs.~\eqref{eq:DimRedBSSNev} exhibit apparently singular terms 
$\sim \tfrac{1}{y}$ or $~\sim \tfrac{1}{y^2}$. While these terms are regular analytically,
numerically the explicit division by $y$ will cause divergences and ``nans'' at and close to $y=0$.
Note, that we encounter similar troublesome terms for the modified Cartoon method 
discussed in Sec.~\ref{ssec:ModCartoon}.

Derive the regularized expressions for these terms, namely
\begin{align}
\label{eq:ExReg}
& \frac{\tg^{iy}\p_{i}X}{y}\,,\quad
  \frac{Y^{y}}{y}\,,\quad
  \frac{\zeta\tg^{yy}-1}{y^2}
\,,\nonumber\\ &
  \frac{1}{y}\left( 2\zeta\tG^{y}{}_{ij} - 2 \delta^{y}{}_{(i}\p_{j)}\zeta \right)
\,,\quad
  \frac{1}{y}\left( \tA^{y}{}_{i} + \delta^{y}{}_{i}\left(\frac{K}{3}-\frac{K_{\zeta}}{\zeta}\right) \right)
\,,
\end{align}
where $X=(\alpha,\chi,\zeta)$ and $Y^{y}=(\beta^{y},\tG^{y})$,
following the example of Eq.~\eqref{eq:Reg1}.

\newpage
\section{Summary and Outlook}\label{sec:Summary}
In the present lecture notes I have introduced the main techniques to explore
time evolutions of higher dimensional BH spacetimes.
I have discussed the key aspects of formulating Einstein's equations as Cauchy problem
for generic spacetime dimension $D$.
Because we can simulate at best $3+1$-dimensional setups with currently avaible
computational resources, we need to re-cast the $(D-1)+1$-dimensional EoMs
as effectively $3+1$-dimensional problems.
This goal can be accomplished by either the Cartoon method or the dimensional reduction by
isometry and I have given a self-consistent introduction to both schemes.
While these methods constrain the phase-space of possible BH configurations to those
with an $SO(D-2)$ or $SO(D-3)$ symmetry, they also have great advantages:
(i) the computational requirements are reduced such that simulations can be performed
efficiently and require only little more resources than ``standard'' numerical evolutions in 
$3+1$ dimensions;
(ii) they allow us to develop a numerical code for generic spacetime dimension $D$,
i.e., there is no need to provide a new implementation for every change of this parameter.

The presented techniques are very powerful tools to investigate 
dynamical spacetimes in $D\geq4$ dimensions and are ready to tackle many open
issues, including 
\begin{itemize}
\item the time evolution of more generic black objects, such as the black ring~\cite{Emparan:2001wn}
and its charged counterpart~\cite{Elvang:2003yy} or
possibly multi-BH solutions, such as the black saturn~\cite{Elvang:2007rd}
and their non-linear stability;
\item the time evolution and stability of charged black holes or black strings;
\item the time evolution of head-on collisions of charged BHs in $D\geq5$, generalizing the study of 
Ref.~\cite{Zilhao:2012gp} in $D=4$;
\item BH spacetimes with AdS asymptotics in $D\geq4$ which are of particular interest for the
gauge/gravity duality~\cite{Maldacena:1997re}. First steps into this direction have been
taken~\cite{Bantilan:2012vu,Chesler:2010bi}
and an ADM-like formulation has been developed~\cite{Heller:2012je,Heller:2011ju} 
but there is still a wide field to explore;
\item a generalization of studies in pure AdS or scalar field-AdS spacetimes which 
have been investigated mainly in spherical symmetry~\cite{Bizon:2011gg,Maliborski:2013jca,Buchel:2013uba}.
\end{itemize}

The proposed possible projects are suited to shed more light 
(i) on the phase-space of higher dimensional BH solutions and their non-linear stability, 
thus complementing perturbative calculations 
(see, e.g., Refs.~\cite{Reall:2012it,Emparan:2008eg,Horowitz:2012nnc} for recent reviews);
(ii) complementary studies regarding the Hoop-conjecture and justifications to model high energy
particle collisions by BHs~\cite{Choptuik:2009ww,East:2012mb, Rezzolla:2012nr};
(iii) on the dynamical evolution of BH-AdS spacetimes and their stability, such as the superradiant
instability for small Kerr-AdS BHs~\cite{Cardoso:2004hs,Hawking:1999dp},
and their counterparts of the CFT side.

\newpage
\section*{Acknowledgements}
I thank the organizers and participants of the NR/HEP2 spring school~\cite{ConfWeb}
for this successful event and many interesting dicussions.
It is a pleasure to thank Hirotada Okawa and Pau Figueras for many fruitful discussions 
and useful comments.
I wish to thank Vitor Cardoso, Leonardo Gualtieri, Carlos Herdeiro, Andrea Nerozzi, Ulrich Sperhake
and Miguel Zilh\~{a}o for our productive collaboration in several projects related to 
this work.

This work has been supported by the {\it ERC-2011-StG 279363--HiDGR} ERC Starting Grant and the STFC GR Roller grant ST/I002006/1.
I also acknowledge support by 
FCT--Portugal through grant nos. SFRH/BD/46061/2008 and CERN/FP/123593/2011
and by {\it DyBHo--256667} ERC Starting Grant
at early stages of this work.

Computations were performed at the cluster "Baltasar-Sete-S\'ois", supported by the {\it{DyBHo--256667}} ERC Starting Grant,
and at the COSMOS supercomputer, part of the DiRAC HPC Facility which is funded by STFC and BIS.
I thankfully acknowledge the computer resources, technical expertise and assistance provided by CENTRA/IST Lisbon 
and the COSMOS support team at the University of Cambridge.


\appendix
\section{Solutions to problems in Sec.~\ref{sec:Decomp}}
\label{asec:app1}
\subsection*{Task~\ref{ssec:ExcPropCurv} -- Properties of the extrinsic curvature}
\begin{enumerate}
\item In order to show that the extrinsic curvature is indeed a spatial quantity we consider its contraction with the 
normal vector. From Eq.~\eqref{eq:def1Kmn} follows
\begin{align}
K_{MN} n^{N} = K_{NM} n^{N}= & n^{N} \gamma^{K}{}_{N} \nabla_{K} n_{M} = 0
\,,
\end{align}
where we use that $\gamma^{K}{}_{N} n^{N}=\delta^{K}{}_{N} n^{N}+n^{K}n_{N}n^{N}=n^{K}-n^{K}=0$.
%
\item In order to derive the relation~\eqref{eq:def2Kmn} between the spatial metric $\gamma_{\bm\bn}$
and extrinsic curvature $K_{\bm\bn}$ we consider the Lie-derivative of the metric along the
vector $u^{M}=\alpha n^{M}$. From Eq.~\eqref{eq:DefLie} we obtain
\begin{align}
\Lie_{\alpha n} \gamma_{MN} = &
   \alpha n^{L} \nabla_{L} \gamma_{MN} + \gamma_{ML}\nabla_{N}(\alpha n^{L}) + \gamma_{NL}\nabla_{M}(\alpha n^{L})
\,\nonumber\\ = &  
   \alpha n^{L} \nabla_{L}(n_{M} n_{N}) + \alpha \gamma_{LM} \nabla_{N} n^{L} + \alpha \gamma_{LN} \nabla_{M} n^{L}
\,\nonumber\\ = &  
   \alpha n_{M} n^{L}\nabla_{L} n_{N} + \alpha n_{N} n^{L}\nabla_{L} n_{M} +\alpha \gamma^{L}{}_{M}\nabla_{N} n_{L} + \alpha \gamma^{L}{}_{N} \nabla_{M} n_{L}
\,\nonumber\\ = &  
   \alpha n_{M} a_{N} + \alpha n_{N} a_{M} - \alpha \gamma^{L}{}_{M} (K_{NL} + n_{N}a_{L}) - \alpha \gamma^{L}{}_{N}(K_{ML} + n_{M} a_{L})
\,\nonumber\\ = &  
   - 2 \alpha K_{MN}
\,,
\end{align}
where we have used Eq.~\eqref{eq:GamG1}, 
the fact that the induced metric is spatial, i.e., $\gamma_{MN} n^{N} = 0$
and the definition of the extrinsic curvature~\eqref{eq:def1Kmn}.
If we now insert Eq.~\eqref{eq:talpbet}, i.e.,
$\alpha n^{M} = t^{M} - \beta^{M}$
we arrive at the desired expression~\eqref{eq:def2Kmn}
\begin{align}
K_{\bm\bn} = & -2\alpha (\p_t -\Lie_{\beta})\gamma_{\bm\bn}
\,.
\end{align}
Note, that I have replaced the spacetime indices $(M,N)$ with the spatial ones $(\bm,\bn)$
because all involved quantities are spatial. 

\end{enumerate}

\subsection*{Task~\ref{ssec:EELam} -- Einstein's equations with cosmological constant}
The Einstein's equation with cosmological constant in the form $E_{MN}=0$ writes
\begin{subequations}
\label{eq:SolEELam}
\begin{align}
E_{MN,1} = & R_{MN}-\frac{1}{2} g_{MN} R + \Lambda g_{MN} - 8\pi G_{D} T_{MN} = 0
\,,\\
E_{MN,2} = & R_{MN} - \frac{2}{D-2}g_{MN}  \Lambda - 8\pi G_{D}\left( T_{MN} - \frac{1}{D-2}g_{MN} T\right) = 0
\,.
\end{align}
\end{subequations}
An example solution for the derivation of the ADM-form of Eqs.~\eqref{eq:SolEELam}
is provided in the \textsc{mathematica} notebook ``GR\_Split\_Sols.nb''
available at Ref.~\cite{ConfWeb}.
For comparison I present the modified equations given by
\begin{subequations}
\label{eq:SolADMLam}
\begin{align}
\H = & [{\rm{ADMflat}}] - 2 \Lambda
\,,\qquad
\M_{\bm} = [{\rm{ADMflat}}]
\,,\\
\p_t\gamma_{\bm\bn} = & [{\rm{ADMflat}}]
\,,\qquad\qquad
\p_t K_{\bm\bn} =  [{\rm{ADMflat}}] - \frac{2}{D-2} \alpha \gamma_{\bm\bn} \Lambda
\end{align}
\end{subequations}
where ``[ADMflat]'' denotes the ADM equations~\eqref{eq:ADMeqs}.

\subsection*{Task~\ref{ssec:EESca} -- Einstein's equations in non-vacuum spacetimes}
The Einstein-Scalar system for a real scalar field $\Phi$ is described by Einstein's equations~\eqref{eq:EEs}
with the energy-momentum tensor 
\begin{align}
T_{MN} = & \nabla_M \Phi \nabla_N \Phi - \frac{1}{2} g_{MN} \nabla^{L}\Phi \nabla_{L}\Phi
\,.
\end{align}
The system is closed by the EoM for the scalar field,
which follows from energy-momentum conservation $\nabla_{M}T^{MN}=0$
and is given by the wave equation
\begin{align}
\S= &\nabla^{M}\nabla_{M} \Phi = 0
\,.
\end{align}
It is useful to introduce the conjugated momentum to the scalar field 
\begin{align}
K_{\Phi} = & - n^{M}\nabla_{M} \Phi = -\Lie_{n}\Phi
\,.
\end{align}
An example solution for the derivation of the ADM-form 
is provided in second part of the \textsc{mathematica} notebook ``GR\_Split\_Sols.nb''
available here~\cite{ConfWeb}.

For comparison, I write down the modified EoMs in the scalar and tensor sector, assuming $D=4,\Lambda=0$:
\begin{align}
\H = & [{\rm{ADMvac}}] - 8\pi G_{D} \left( K_{\Phi}^2 + D^{i}\Phi D_{i}\Phi \right)
\,,\\ 
\M_{i} = & [{\rm{ADMvac}}] - 8 \pi G_{D} K_{\Phi} D_i \Phi
\,,\\
\p_t\gamma_{ij} = & [{\rm{ADMvac}}]
\,,\\
\p_t K_{ij} = & [{\rm{ADMvac}}]  - 8\pi G_{D} \alpha D_i\Phi D_{j}\Phi
\,,\\
\p_t \Phi =  &- \alpha K_{\Phi} + \Lie_{\beta} \Phi
\,,\\
\p_t K_{\Phi} = & - \alpha\left( D^{i}D_{i} \Phi - K K_{\Phi} \right) - D^{i}\alpha D_{i}\Phi + \Lie_{\beta}K_{\Phi}
\,,
\end{align}
where ``[ADMvac]'' denotes the vacuum ADM equations.

\section{Solutions to problems in Sec.~\ref{sec:Cartoon}}
\label{asec:app2}
\subsection*{Task~\ref{sssec:Cartoon2x} -- 
Cartoon method in $D=5$ with $U(1)\times U(1)$ symmetry}
In order to develop the Cartoon method for the doubly spinning Myers-Perry BH in $D=5$
we will adapt our ``recipe'' introduced in Sec.~\ref{ssec:Cartoon5D}.
The key idea is to apply the Cartoon method twice.
This \textit{double Cartoon method} has originally been presented in Ref.~\cite{Yoshino:2009xp}.
\begin{enumerate}
\item Let us denote the spatial Cartesian coordinates as $(x,y,z,w)$. 
The doubly spinning Myers-Perry BH in $D=5$ is a spacetime with a $U(1)\times U(1)$
symmetry. If we choose the $x-y$- and $z-w$-planes as planes of symmetry,
the spacetime exhibits two KV fields
$\p_{\varphi}=x\p_y - y\p_x$ and $\p_{\psi}=z\p_{w}-w\p_{z}$.
The Cartesian coordinates are related to 
polar-like coordinates $(\rho_{1},\varphi,\rho_{2},\psi$) by
\begin{align}
& x=\rho_{1} \cos\varphi\,,\quad
  y=\rho_{1} \sin\varphi\,,\quad
  \text{with}\quad
  \rho_{1} = \sqrt{x^2+y^2} 
\,,\nonumber\\
& z=\rho_{2} \cos\psi\,,\quad
  w=\rho_{2} \sin\psi\,,\quad
 \text{with}\quad
  \rho_{2} = \sqrt{z^2+w^2}
\,;
\end{align}
\item We now have to perform two rotations with the linear maps
$R_{1}:\,^{(4)}\Sigma_t \rightarrow \,^{(4)}\Sigma_t$ and
$R_{2}:\,^{(4)}\Sigma_t \rightarrow \,^{(4)}\Sigma_t$ given by
\begin{align}
\label{eq:Rot5DU1U1}
\left( R_{1}(\varphi_0)^{\bm}{}_{\bn} \right) = &
\left( \begin{matrix}
\cos\varphi_{0} & -\sin\varphi_{0} & 0 & 0 \\
\sin\varphi_{0} &  \cos\varphi_{0} & 0 & 0 \\
0 & 0 & 1 & 0 \\
0 & 0 & 0 & 1 
\end{matrix} \right)
\,,\,\,\,
\left( R_{2}(\psi_0)^{\bm}{}_{\bn} \right) = &
\left( \begin{matrix}
1 & 0 & 0 & 0 \\
0 & 1 & 0 & 0 \\
0 & 0 & \cos\psi_0 & -\sin\psi_0 \\
0 & 0 & \sin\psi_0 &  \cos\psi_0 
\end{matrix} \right)
\,,\quad
%
\end{align}
where $(\varphi_0,\psi_0)$ are the  rotation angles;
\item First, we compute the numerical data at the grid points\\ 
      $(x,y=0,z,w=0)$;
\item To employ the Cartoon method for the first time 
      we have to interpolate the grid functions to points $p_1=(\rho_1,\varphi=0,z,0)$.
      Then, we generate function values at grid points $p'_1=(x,y\neq0,z,0)$ by rotating
      the data around the angle $\varphi_0$;
\item In order to generate the final data we have to apply the Cartoon method a second
      time. Therefore we have to interpolate function values from\\ 
      $p'_1=(x,y\neq0,z,0)$ to $p_2=(x,y,\rho_2,\psi=0)$.
      Finally, we rotate the data by an angle $\psi_0$ from $p_2$ to $p'_2=(x,y,z,w\neq0)$.
      Analogous to Eq.~\eqref{eq:Trafo5DU1} scalar, vector and 2-tensors transform as
\begin{subequations}
\label{eq:Trafo5DU1U1}
\begin{align}
\Psi(x,y,z,w)   = & \Psi(\rho_1,0,\rho_2,0) \,,\\
V^{\bm}(x,y,z,w)  = & R_{2}^{\bm}{}_{\bn} R_{1}^{\bn}{}_{\bk} V^{\bk}(\rho_1,0,\rho_2,0) \,,\\
S_{\bm\bn}(x,y,z,w) = & (R_{2}^{-1})^{\bk}{}_{\bm} (R_{2}^{-1})^{\bl}{}_{\bn} 
                        (R_{1}^{-1})^{\bm}{}_{\bp} (R_{1}^{-1})^{\bn}{}_{\bq}
                        S_{\bp\bq}(\rho_1,0,\rho_2,0) 
\,.
\end{align}
\end{subequations}
\end{enumerate}
%

\newpage
\subsection*{Task~\ref{sssec:Cartoon6D} -- Cartoon method in $D\geq6$} 
I will show the verification of the symmetry relations~\eqref{eq:CartoonDsubS}--~\eqref{eq:CartoonDsubT} 
exemplarily for the conformal factor $\chi=\gamma^{-\tfrac{1}{3}}$ and 
the conformal metric $\tg_{ij}=\chi \gamma_{ij}$. Both quantities are tensor densities 
of weight $\mathcal{W}=-\tfrac{2}{D-1}=-\tfrac{2}{3}$, where we use the fact that $D=4$ for the base space.

\noindent{{\bf{Verification of Eq.~\eqref{eq:CartoonDsubS}:}}}
As a first step I verify these relations for the determinant $\gamma$ of the physical metric,
which I will use in the following. $\gamma$ is a tensor density of weight $\mathcal{W}=2$.
Thus, we obtain
\begin{subequations}
\label{eq:CartoonDsubGam}
\begin{align}
\p_{\bmu}\gamma = & \gamma \gamma^{ij}\p_{\bmu}\gamma_{ij} = 0
\,,\\
\p_{\bmu}\p_{\bnu}\gamma = & \gamma \gamma^{ij}\p_{\bmu}\p_{\bnu} \gamma_{ij}
        + \gamma^{kl}\p_{\bmu}\gamma \p_{\bnu}\gamma_{kl}
        - \gamma \gamma^{ik}\gamma^{jl}\p_{\bmu}\gamma_{ij} \p_{\bnu}\gamma_{kl}
        = \gamma \gamma^{ij}\p_{\bmu}\p_{\bnu} \gamma_{ij}
        = \delta_{\bmu\bnu} \frac{\p_{z}\gamma}{z}
\,,
\end{align}
\end{subequations}
using Eqs.~\eqref{eq:CartoonDsubT} for the physical metric.
The conformal factor is related to the determinant $\gamma$ of the physical metric 
by $\chi=\gamma^{-\frac{1}{3}}$. We have just verified that the relations~\eqref{eq:CartoonDsubS} are valid 
for $\gamma$. Using this fact, we can show that
\begin{subequations}
\begin{align}
\p_{\bmu}\chi = & -\frac{1}{3}\gamma^{-\frac{4}{3}}\p_{\bmu}\gamma = 0
\,,\\
\p_{\bmu}\p_{\bnu} \chi = & \frac{4}{9}\gamma^{-\frac{7}{3}}\p_{\bmu}\gamma\p_{\bnu}\gamma 
                           -\frac{1}{3}\gamma^{-\frac{4}{3}}\p_{\bmu}\p_{\bnu}\gamma
        = -\frac{1}{3}\gamma^{-\frac{4}{3}}\delta_{\bmu\bnu}\frac{\p_{z}\gamma}{z}
        = \delta_{\bmu\bnu}\frac{\p_{z}\chi}{z}  
\,.
\end{align}
\end{subequations}
\noindent{{\bf{Verification of Eqs.~\eqref{eq:CartoonDsubT}:}}}
As an example, I will show the derivation of the symmetry relation~\eqref{eq:CartoonDsubT2} for 
the conformal metric. 
From $\tg_{ij}=\chi\gamma_{ij}=\gamma^{-\frac{1}{3}}\gamma_{ij}$ follows
\begin{align}
\p_{\bmu}\p_{\bnu} \tg_{ij} = & \gamma^{-\frac{1}{3}}\p_{\bmu}\p_{\bnu}\gamma_{ij} - \frac{1}{3}\gamma^{-\frac{4}{3}}\gamma_{ij} \p_{\bmu}\p_{\bnu}\gamma
\nonumber\\ 
        = & - \frac{1}{3}\gamma^{-\frac{4}{3}}\gamma_{ij} \delta_{\bmu\bnu} \frac{\p_{z}\gamma}{z}
          + \gamma^{-\frac{1}{3}}\delta_{\bmu\bnu} \left( \frac{\p_{z}\gamma_{ij}}{z} -\delta_{iz}\frac{\gamma_{ij}}{z^2} + \delta_{iz}\delta_{jz} \frac{2\gamma_{ww}-\gamma_{zz}}{z^2} \right)
\nonumber\\ 
        = & \frac{\p_{z}\tg_{ij}}{z} -\delta_{iz}\frac{\tg_{ij}}{z^2} + \delta_{iz}\delta_{jz} \frac{2\tg_{ww}-\tg_{zz}}{z^2}
\,,
\end{align}
where we have used Eqs.~\eqref{eq:CartoonDsubGam} and~\eqref{eq:CartoonDsubT} for $\gamma$ and $\gamma_{ij}$.
In a similar manner one can verify the remaining relations ~\eqref{eq:CartoonDsubS}--~\eqref{eq:CartoonDsubT} 
also for the conformal, trace-free part of the extrinsic curvature $\tA_{ij}=\chi A_{ij}$.
Using Eq.~\eqref{eq:densGam} we can also derive the expressions for to the conformal connection function $\tG^{i}$.

\section{Solutions to problems in Sec.~\ref{sec:DimRed}}
\label{asec:app3}
\subsection*{Task~\ref{ssec:DimRedADM} -- ADM form of the modified Einstein's equations}
When performing the dimensional reduction by isometry with the assumed symmetries
we obtain Einstein's equations in $D=4$ coupled non-minimally to a scalar field.
The EoMs are given in Eqs.~\eqref{eq:DimRedEoMs}.
An example solution to derive the ADM form~\eqref{eq:DimRedADMs},~\eqref{eq:DimRedADMg} and~\eqref{eq:DimRedADMc} 
of the EoMs is given in the \textsc{mathematica} notebook ``GR\_Split\_Sols.nb''
available online~\cite{ConfWeb}.

\subsection*{Task~\ref{ssec:Regularization} -- Regularization of terms $\sim \tfrac{1}{y}$ and $~\sim \tfrac{1}{y^2}$}
The BSSN evolution Eqs.~\eqref{eq:DimRedBSSNev} obtained from the dimensional reduction by isometry
as well as those of the modified Cartoon method
contain terms which are apparently singular along one axis.
Analytically these terms are regular, but the explicit division by $0$ would still cause
problems numerically. Therefore, we have to regularize these terms and 
explicitely substitute them close to the axis.
For the sake of discussion let us focus on singular terms 
of Eqs.~\eqref{eq:DimRedBSSNev} and~\eqref{eq:HDFgBSSNcoupling}.
An analogous procedure will yield regularized expressions for the  
modified Cartoon method (see Eqs.~\eqref{eq:CartoonDsubS}--~\eqref{eq:CartoonDsubT}).

\begin{enumerate}
\item 
Let us start the discussion with the term $\frac{Y^{y}}{y}$ with $Y^{y}\in(\beta^{y},\tG^{y})$.
The functions $Y^{y}$ are linear in $y$ near the axis
and, thus, their Taylor expansion around $y=0$ is given by 
$Y^{y} = y \p_{y} Y^{y}|_{y=0} + \O(y^2) = y Y^{y}_{1} + \O(z^2)$.
If we take the limit $y\rightarrow0$ we obtain
\begin{align}
\label{eq:RegY}
\lim_{y\rightarrow0} \frac{Y^{y}}{y} = & Y^{y}_{1} = \p_y Y^{y}|_{y=0}
\,.
\end{align}
\item
Next, let us consider the terms $\frac{\tg^{iy}\p_{i}X}{y}$ in Eq.~\eqref{eq:ExReg} where
$X\in(\alpha,\chi,\zeta)$.
The derivative of the scalar (densities) behave similar as a vector and together 
with the symmetry relations of the conformal metric we get
\begin{align}
\lim_{y\rightarrow0} \left( \frac{\tilde{\gamma}^{yi}}{y}\p_{i} X \right) = &
        \p_{y} \tg^{ay}\p_{a} X + \tg^{yy} \p_{y}\p_{y} X
\,,
\end{align}
where $a\in(x,z)$.
\item
The regularity of $\frac{\zeta\tg^{yy}-1}{y^2}$ follows from the 
requirement that there should be no conical singularities at the axis.
This requirement translates into $\zeta\tg^{yy} = 1+\O(y^2)$ and it follows
\begin{align}
\lim_{y\rightarrow0} \frac{\zeta\tg^{yy}-1}{y^2} = & 
        \frac{1}{2}\left( \zeta \p_{y}\p_{y} \tg^{yy} + \tg^{yy} \p_{y}\p_{y} \zeta \right)
\,.
\end{align}
\item
Next, let us focus on the term $\frac{1}{y}\left( 2\zeta\tG^{y}{}_{ij} - 2 \delta^{y}{}_{(i}\p_{j)}\zeta \right)$.
We expand the Christoffel symbols and apply Eq.~\eqref{eq:RegY} as well as the fact that
there should not be any conical singularity at the axis.
Then, the regularized expressions become
\begin{align}
\lim_{y\rightarrow0}\frac{\p_{y}\zeta - \zeta \tG^{y}{}_{yy}}{y} = &
        \p_y\p_y\zeta - \frac{\zeta}{2}\tg^{yy}\p_y\p_y\tg_{yy} 
        + \zeta \p_{y}\tg^{ay} \left(\frac{1}{2}\p_{a}\tg_{yy} - \p_{y}\tg_{ay}\right)
\,,\nonumber\\
\lim_{y\rightarrow0}\frac{\p_{a}\zeta - 2\zeta\tG^{y}{}_{ay}}{y} = & 0
\,,\\
\lim_{y\rightarrow0}\frac{2\zeta\tG^{y}{}_{ab}}{y} = &
        \zeta\tg^{yy}\left( 2\p_{y}\p_{(a}\tg_{b)y} - \p_{y}\p_{y}\tg_{ab} \right)
        + \zeta \p_{y}\tg^{yc}\left( \p_{a}\tg_{bc} + \p_{b}\tg_{ac} - \p_{c}\tg_{ab} \right)
\,,\nonumber
\end{align}
where $(a,b,c)\in(x,z)$ only.

\item 
Finally, let us regularize the term
$ \frac{1}{y}\left( \tA^{y}{}_{i} + \delta^{y}{}_{i}\left(\frac{K}{3}-\frac{K_{\zeta}}{\zeta}\right) \right)$,
where I will distinguish between the cases $i=y$ and $i=a\neq y$.
The latter case results in
\begin{align}
\lim_{y\rightarrow0}\frac{\tA^{y}{}_{a}}{y} = & 
          \lim_{y\rightarrow0}\frac{\tg^{iy}\tA_{ai}}{y}
        = \tA_{ab}\p_{y}\tg^{by} + \tg^{yy}\p_{y}\tA_{ay}
\,.
\end{align}
In order to compute the case $i=y$ let us consider the time derivative of the term
$\zeta-\tg_{yy}$. Let us first note that together with $\lim_{y\rightarrow0}\tg^{yy}=1/\tg_{yy} + \O(y^2)$
and the condition $\lim_{y\rightarrow0}\zeta\tg^{yy} = 1+\O(y^2)$
we obtain 
\begin{align}
\lim_{y\rightarrow0}(\zeta-\tg_{yy}) = & \O(y^2)
\,.
\end{align}
Then, its time derivative becomes
\begin{align}
\O(y^2) =\lim_{y\rightarrow0}\p_t(\zeta-\tg_{yy}) = & 
        -2\alpha\zeta\left(\tg^{iy}\tA_{iy} + \frac{K}{3} - \frac{K_{\zeta}}{\zeta} \right) + \O(y^2)
\,,
\end{align}
which implies
\begin{align}
\lim_{y\rightarrow0}\frac{1}{y}\left( \tg^{iy}\tA_{iy} + \frac{K}{3}-\frac{K_{\zeta}}{\zeta} \right)
= & 0
\,.
\end{align}

\end{enumerate}

\clearpage

\bibliographystyle{h-physrev4}
\bibliography{nrhep}

\end{document}